\documentclass[useAMS,usenatbib]{mn2e}
\usepackage{setspace}

\RequirePackage{epsfig}
\RequirePackage{subfigure}

\def\imcat{IMCAT}
\def\shlets{Shapelets}

\def\rar{\mbox{$\rightarrow$}}

\def\sigmav{\mbox{$\sigma_v$}}

\def\eqref{eq.~\ref}
\def\figref{Fig.~\ref}
\def\secref{Sec.~\ref}
\def\tblref{Table~\ref}

\newcommand{\bi}{\begin{itemize}}
\newcommand{\ei}{\end{itemize}}

\newcommand{\benum}{\begin{enumerate}}
\newcommand{\eenum}{\end{enumerate}}

\def\ellipy{ellipticity}

\def\ellips{ellipticities}

\def\mag{magnitude}
\def\Mag{Magnitude}

\def\avg{average}

\def\sext{SExtractor}

\def\interpln{interpolation}

\def\param{parameter}

\def\bea{\begin{eqnarray}}
\def\eea{\end{eqnarray}}

\def\be{\begin{equation}}
\def\ee{\end{equation}}

\def\rmd{\rm{d}}

\def\eone{\mbox{$e_1$}}
\def\etwo{\mbox{$e_2$}}

\def\rp{\right)}
\def\lp{\left(}

\def\bkgd{background}
\def\Bkgd{Background}

\def\Dls{D_{ls}}

\def\chisq{\mbox{$\chi^2$}}

\def\kmsec{ \, {\rm km/sec} }

\def\corresp{corresponding}

\def\ssa{single sheet approximation}

\def\dof{{\rm dof}}

\title[An Analysis of DES Cluster Simulations through the IMCAT and Shapelets Weak Lensing Pipelines]{An Analysis of DES Cluster Simulations through the IMCAT and Shapelets Weak Lensing Pipelines}

\author[Gill et al.]{M.S.S. Gill$^{1}$\thanks{Email: msgill@astronomy.ohio-state.edu}, J.C. Young$^{1}$,
J.P. Draskovic$^{1}$, K. Honscheid$^{1}$, H. Lin$^{2}$, N. Kuropatkin$^{2}$,  \cr
P. Martini$^{1,3}$, M.Peeples$^{3}$, E. Rozo$^{1,4}$, G.N. Smith$^{1}$, D.H. Weinberg$^{1,3}$ \\
\vspace*{-6pt} {\small \em$^{1}$ Center for Cosmology and AstroParticle Physics, The Ohio State University, Columbus, Ohio 43210, USA}\\
\vspace*{-6pt} {\small \em$^{2}$ The Fermi National Accelerator Laboratory, Batavia, Illinois 60510, USA} \\
\vspace*{-6pt} {\small \em$^{3}$ Department of Astronomy, The Ohio State University, Columbus, Ohio 43210, USA}\\
\vspace*{-6pt} {\small \em$^{4}$ The Kavli Institute for Cosmological Physics, 5640 South Ellis Avenue, Chicago, Illinois  60637, USA }}

\begin{document}

\pagerange{\pageref{firstpage}--\pageref{lastpage}} \pubyear{2009}

\maketitle

\label{firstpage}

\begin{abstract}  
  We have run two completely independent weak lensing analysis
  pipelines on a set of realistic simulated images of a massive galaxy
  cluster with a singular isothermal sphere profile (galaxy velocity
  dispersion $\sigma_v=1250\ \kmsec$). The suite of images was
  constructed using the simulation tools developed by the Dark Energy
  Survey.  We find that both weak lensing pipelines can accurately
  recover the velocity dispersion of our simulated clusters,
  suggesting that current weak lensing tools are accurate enough for
  measuring the shear profile of massive clusters in upcoming large
  photometric surveys.  We also demonstrate how choices of some cuts
  influence the final shear profile and \sigmav\ measurement.
  Analogously to the STEP program, we make all of these cluster
  simulation images publically available for other groups to analyze
  through their own weak lensing pipelines.
\end{abstract}

\begin{keywords}
Cluster Weak Lensing  -- Stage III Photometric Surveys -- Dark Energy Survey 
\end{keywords}

% Sec 1
\section{Introduction}

Gravitational weak lensing, by both galaxy clusters and large scale
structure as cosmic shear, has become one of the most promising
observational avenues in recent times to place constraints on
cosmological parameters and models (e.g.  \citealt{detf},
\citealt{peacock}, \citealt{abbott}).  This is partly because it does
not suffer from the inherent uncertainties involved in extrapolating
cluster masses from other measurement techniques such as galaxy
population richness class, x-ray luminosity of the hot intracluster
gas, or galaxy velocity dispersion.  In some sense, cluster weak
lensing is the most direct measure of the total mass of a cluster,
both seen and unseen. It is thus highly critical to ascertain exactly
how accurate and precise the results from available current pipelines
are in both the cluster and cosmic shear contexts.  The cosmic shear
context has especially been tested in recent years through the STEP
and GREAT collaborations (\citealt{massey}, \citealt{heymans},
\citealt{great}), but there has been no analogous collaborative effort
to verify the validity of lensing pipelines in the context of cluster
weak lensing.  With the public release of the images used in this
analysis, we aim to begin rectifying this in anticipation of upcoming
major Stage III and Stage IV surveys \citep{detf}, which will collect
very large amounts of data useful for cluster weak lensing.

In the case of the STEP and GREAT programs, sets of simulated images
were made with the aim especially of comparing and optimizing
pipelines designed for the measurement of cosmic shear.  For this
purpose, having images with constant shear and constant Point Spread
Function (PSF) across the field was sufficient for verifying several
of the primary properties of the pipelines. However, for cluster weak
lensing analyses we know that the shear will vary both in magnitude
and direction across the field.  Thus to test pipelines for
applicability to cluster weak lensing analyses, it is more suitable
and realistic to make simulated images which incorporate variation of
both the shear and the PSF across the field. We did this by making
simulated images in which an isothermal sphere shear is applied across
the entire focal plane, and which further incorporate a PSF that
varies linearly in anisotropy size and magnitude identically across
each CCD.

One example of a major upcoming Stage III observational campaign is
the Dark Energy Survey (DES) \citep{abbott}, scheduled to begin taking
data in the Fall of 2011.  DES will use four main techniques to place
constraints on Dark Energy: (1) galaxy cluster number count
distributions, (2) weak lensing cosmic shear measurements, (3) galaxy
angular clustering measurements, and (4) distance measurements to Type
Ia supernovae.  Cluster weak lensing to determine cluster masses, and
other properties such as concentrations, is also one of the integral
components of the DES program, and will play an important role in the
calibration of the first technique in the above list.  

As part of the preparation for DES, weak lensing pipelines are being
thoroughly tested for accuracy on simulations before the survey
begins.  In addition, the DES data management pipeline and analysis tools
are being extensively exercised in pre-observation DES Data Challenges
with realistic images. Following from these, and using the tools
developed for the Data Challenges, we constructed a set of realistic
images containing a simulated galaxy cluster.  We then processed these
images through two completely independent weak lensing pipelines and
demonstrated that the shear measurements are robust enough to
reconstruct the cluster mass reliably. We also studied how different
configuration and selection choices in the weak lensing pipelines
affect the final cluster mass reconstruction.

We analyzed these sample cluster images with two pipelines based on
publically available weak lensing codes: \imcat\ -- an implementation
of the Kaiser, Squires, \& Broadhurst \citep{ksb} method -- and
Shapelets \citep{refregier}, and we present the results in this paper.
It will be very useful to test multiple weak lensing pipelines on
these cluster simulations, and we have made them publically available
to the wider cosmological and weak lensing communities for this
purpose at the following website: {\bf
  [http://ccapp.osu.edu/DEScluster]}.

The structure of the paper is as follows: in Section~\ref{sec:cluster}
we describe the simulated cluster.  In Section~\ref{sec:pipeline} we
describe the overall pipeline flow, object extraction and processing
as well as test our PSF removal procedures on the simulated cluster.
In Section~\ref{sec:results} we give our results, in
Section~\ref{sec:varstud} we show how the results vary with certain
cuts, and in Section~\ref{sec:conclusions} we give our conclusions.

% Sec 2
\section{Simulated Cluster Image Properties}
\label{sec:cluster}

\subsection{Description of Background Galaxy Properties}
\label{sec:simlndescrip}

Several cluster images were created with the gravitational weak
lensing shear determined by the Singular Isothermal Sphere (SIS)
model using the tools developed by the DES simulation group.  Objects
in the simulated images came from the same simulated galaxy and
stellar catalogs used to populate image simulations generated for the
annual DES Data Challenge process, which is carried out to help
develop and test DES data processing and science analysis pipelines.
Specifically, the catalogs used in this paper were the ones employed
for the so-called DES ``Data Challenge 4,'' carried out in 2008-2009.

The galaxy catalogs used the publically available Hubble Volume
\citep{colberg} as the parent dark matter N-body simulation box.
Specifically the $\Lambda$CDM Hubble Volume simulation was used, with
cosmology $\Omega_M = 0.3, \Omega_\Lambda = 0.7, h=0.7, \sigma_8 =
0.9$ and a light cone with redshift limit $z = 1.4$.  Galaxies were
assigned to Hubble Volume dark matter particles using the ADDGALS
method (see Appendix A of \citealt{gerdes}, \citealt{wechsler},
\citealt{wechsler2}), whereby each galaxy was assigned the properties
of a real Sloan Digital Sky Survey (SDSS) galaxy, and the procedure
reproduces the local luminosity-color-density correlations from the
SDSS.  ADDGALS also reproduces the magnitude-dependent 2-point
function, which is important to get correct cluster color-magnitude
diagrams.  In particular, to populate the galaxies in the image we
used the Data Challenge 4 DES mock galaxy catalog, which covers a 573
deg$^2$ subset of the Hubble Volume sky area.  This galaxy catalog has
a surface density of about 13 galaxies per arcmin$^2$ down to an
apparent magnitude limit in the $i$ band of about $i=24$.  The
luminosity function model used was that of \citet{blanton} for the
SDSS, with simple passive luminosity evolution applied ($M_*$ brightens
by 1.3 mag per unit redshift) irrespective of galaxy type.  No
redshift evolution beyond the local SDSS color-environment
correlations was included.  Galaxies were assigned shape parameters
using shapelet coefficients \citep{refregier}, where the master
shapelet coefficient distribution used was derived from a set of about
18,000 galaxies in $i$-band CFHT MegaCam images, taken from the public
Cosmic Evolution Survey (COSMOS; \citet{scoville}) data set.  The
shapelet decompositions were carried out to order n=15, corresponding
to using a total of 136 shapelet coefficients to describe each galaxy.
The COSMOS/CFHT galaxies were assigned to the DES mock catalog
galaxies using a simple nearest neighbor matching in the space of
$griz$ magnitudes.  No size or shape evolution was included other than
redshift effects due to cosmology.

The stellar catalogs were based on real USNO-B stars \citep{monet} at
the bright end ($r < 20$) and simulated stars at the faint end ($r >
20$), with the latter derived from a web based tool employing the
Besancon stellar population synthesis model for our
Galaxy\footnote{http://model.obs-besancon.fr/} to generate
simulated stars.

\subsection{Details of Cluster Simulation}
\label{sec:clusdescrip}

The simulated cluster is located at a redshift of 0.33, with $M_{200}
= 1.5\cdot 10^{15} M_{\odot}$, $r_{200}=1$ Mpc and $\sigmav = 1250$ km/sec, where $M_{200}$
($r_{200}$) is the mass (radius) out to a distance within which the
average overdensity is 200 times the critical density of the
universe at that redshift and \sigmav\ is the velocity dispersion for
an SIS profile (see Sec.~\ref{sec:fitsis} and for details on SIS
properties see e.g.  Sec.~3.1 of \citet{narayanBartelmann} .  In
order to avoid the strong lensing regime within the Einstein radius,
which for this cluster for sources at $z=1$ is about 27\arcsec 
(see Sec.~\ref{sec:fitsis}), the shear is set to zero for objects
within a certain radius.  
% Sources at $z=1$ have a $\beta \equiv {D_{ls} \over D_s}$ factor of about 0.6.
To be conservative we choose the value of this radius by excluding the
region where objects would have tangential shear $\gamma_t > 0.2$ in
the limit of $z \rar \infty$, which corresponds to the area within a
radius of about 100'' from the cluster center.  Using the pixel scale
of the Dark Energy Camera (DECam) of 0.27 arcseconds per pixel
\citep{honscheid}, this corresponds to being within about 400 pixels
from the cluster center.  Except for this region, galaxy objects have
SIS shear assigned accordingly across the full 3 square degree image
mosaic.  To get some sense of the scale of the shear variation, for a
source at $z=1$ the shear falls to about $\gamma = 0.02$, at a radius
of 20\arcmin (6 Mpc) from the cluster center, roughly one third of the
image plane radius.

\begin{figure}
\begin{center}
\subfigure[Original file -- no noise]{\epsfig{file=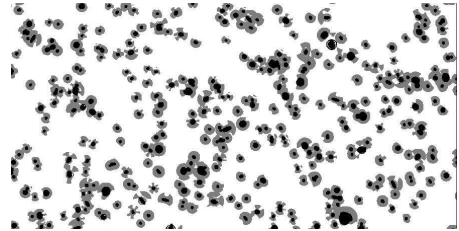,height=3cm,angle=0} } \\
\subfigure[ Low noise file - without foreground galaxies ]{ \epsfig{file=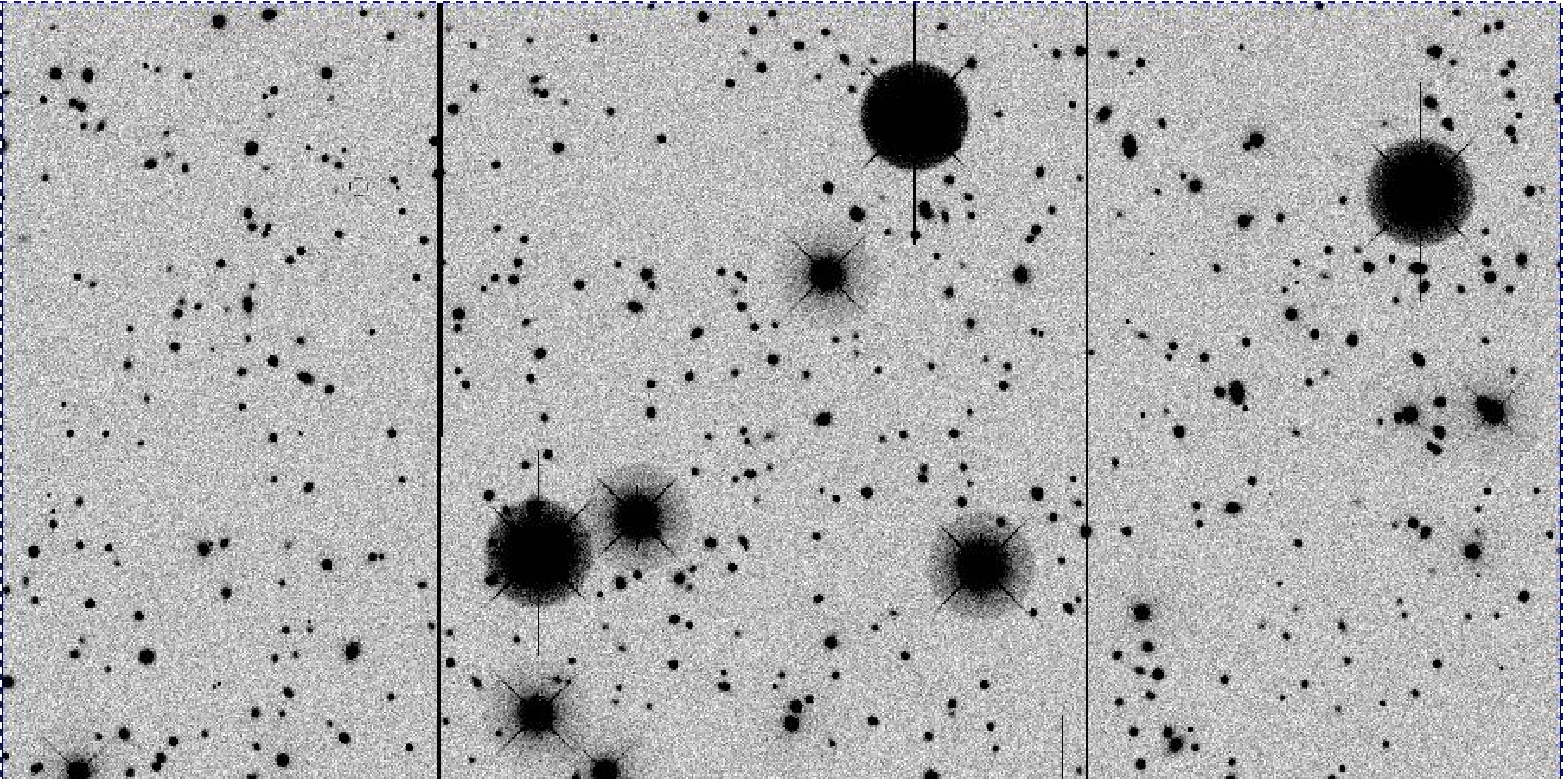,height=3cm,angle=0} }\\
\subfigure[ Low noise file - with foreground galaxies ]{ \epsfig{file=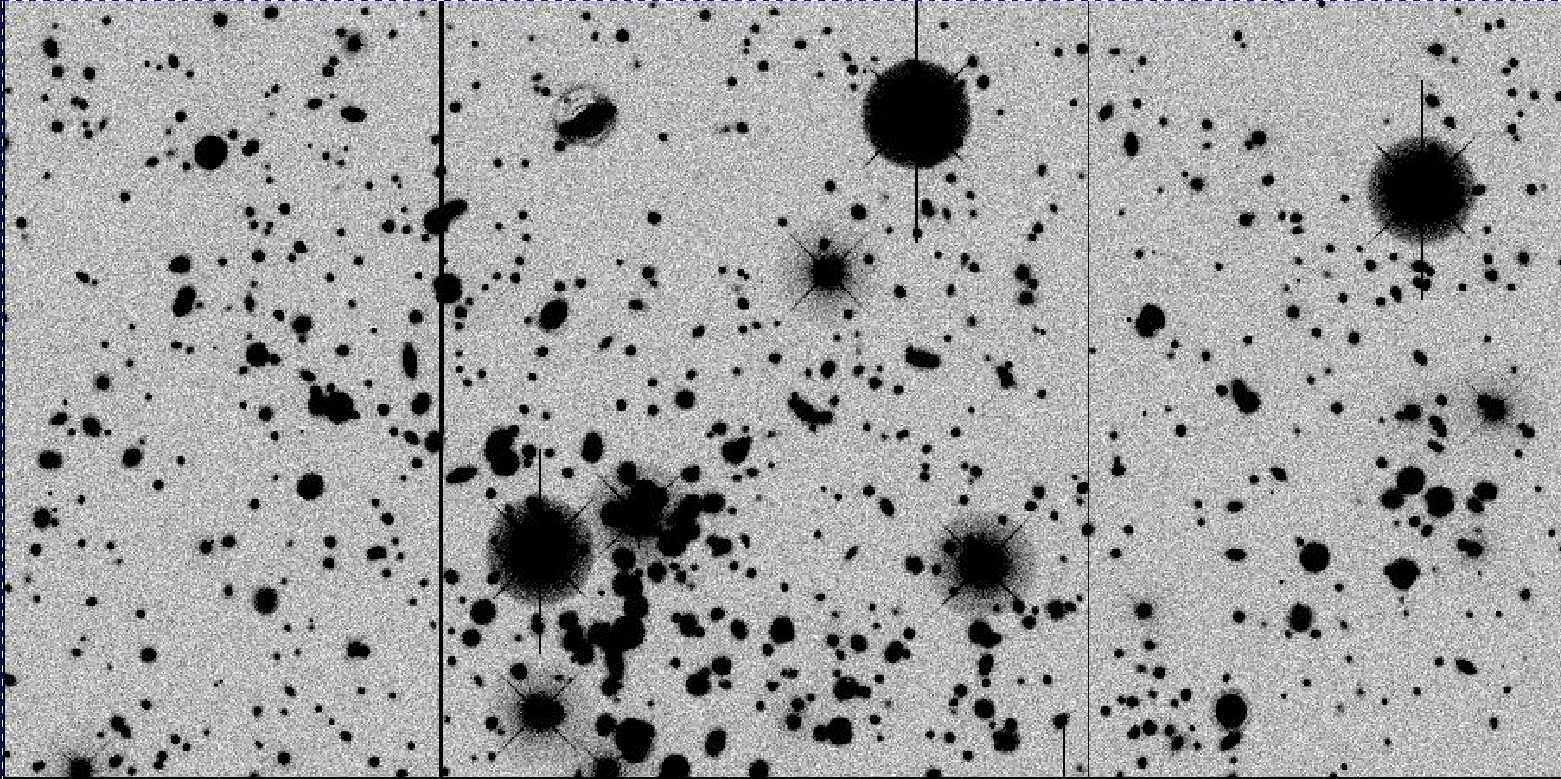,height=3cm,angle=0} }
\subfigure[ High noise file - without foreground galaxies ]{ \epsfig{file=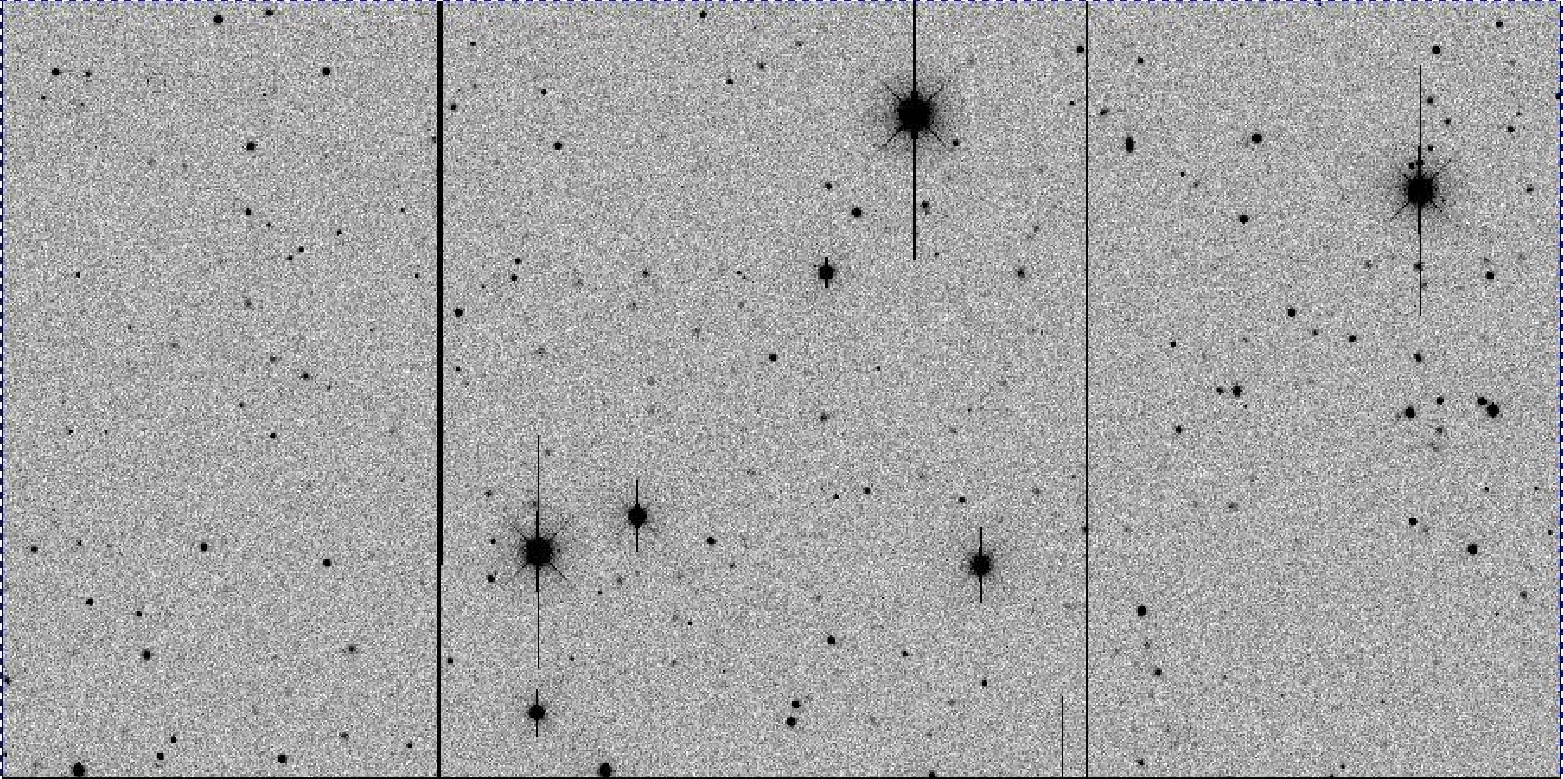,height=3cm,angle=0} }
\subfigure[ High noise file - with  foreground galaxies ]{ \epsfig{file=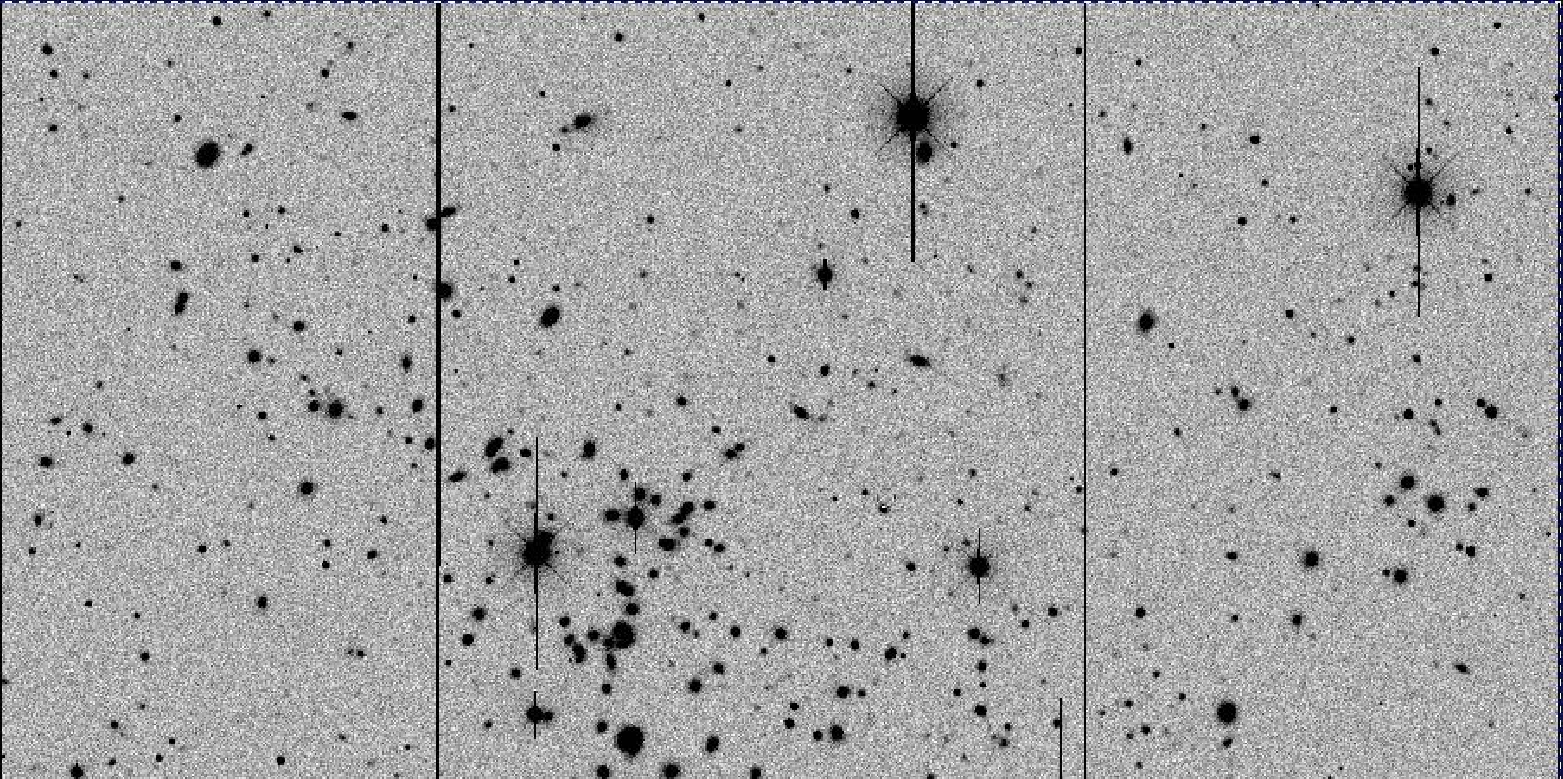,height=3cm,angle=0} } 
\end{center}
\caption{Partial focal plane images from the simulated cluster, for the portion containing the cluster, 
  with size about 1000 pixels high, and 2000 pixels across. Background
  galaxies are visible as well as clear diffraction spikes (from the
  DECam telescope assembly) on the saturated foreground stars in the
  four images that contain them, and foreground cluster galaxies in
  the two images that contain them.  It is clear that many more
  objects are seen in the low noise files.  The vertical lines running
  from top to bottom in all four images with noise are bad CCD
  columns. }
\label{fig:simclust}
\end{figure}

Each galaxy object generated according to this prescription has five
magnitudes in the SDSS-like filters $g,r,i,z$ and $y$, as well as an
assigned redshift, which are all contained in truth catalog files.
Seven image files were produced with varying levels of complexity, and
their properties are summarized in Table~\ref{table:simimage}.
Partial images for five of the files are shown in
Fig.~\ref{fig:simclust}.  In order to test how the pipelines vary for
differing noise levels, there are three levels of noise included in
the simulated images: No Noise, LN (Low Noise) and HN (High Noise).
The LN images are what would be seen by DES with a 600 s exposure with
readout and object photon noise included but without sky noise. The
effects of the sky noise are present in the HN images which have a 13
times higher noise level. Our HN images should be fairly similar to
nominal DES exposures which will be 100 seconds long with 5 passes
throughout the survey.

\begin{table*}
  \caption{Brief description of the simulated images.}
  \begin{tabular}{@{}|l|l|l|l|l|l|l|l|l| @{}}
  \hline
  \hline
 File Description & Noise & Shear & Included Objects \\ 
\hline
\hline
 Original Non-sheared  & None &  None  &  Background Galaxies \\
\hline
 Sheared   & None & Cluster Shear &  Background Galaxies  \\
\hline
 High Noise  &  High  &  Cluster Shear &   Background Galaxies   \\
\hline
 High Noise + PSF Applied  &  High  &  Cluster Shear &   Background Galaxies   \\    
                 &        &   + PSF         &   Stars               \\
\hline
High Noise  + PSF Applied + Foreground Galaxies  & High  &  Cluster Shear &   Background Galaxies    \\
                                &       &  + PSF         &   Stars                \\
                                &       &                &  Foreground Galaxies  \\
\hline
 Low Noise + PSF Applied  &  Low            &  Cluster Shear &  Background Galaxies  \\
                      &                 &  + PSF         &  Stars                \\
\hline
 Low Noise + PSF Applied + Foreground Galaxies   &  Low   & Cluster Shear &   Background Galaxies  \\
                           &        & + PSF         &  Stars                \\
                                     &        &               &  Foreground Galaxies  \\
\hline
\hline
\end{tabular}
 \label{table:simimage}
\end{table*}

\begin{table*}
\caption{Rounded initial numbers of objects in the labelled column categories of \tblref{table:simimage}, all in thousands.  No cut has been applied to the files with foreground galaxies, so all of these are included in the yield for Galaxy Objects.  100k galaxies across the DES focal plane would correspond to about 9 galaxies per square arcminute.}
\begin{tabular}{|l|l|l|l|l|l|}
\hline
\hline
File & Raw Objects & Cleaned Objects & Stellar Objects & Galaxy Objects  \\
& (thousands) & (thousands) & (thousands) & (thousands) \\
\hline
\hline
 Original & 105 & 28 & n/a & 28  \\
\hline
 Sheared & 105 & 28  & n/a & 28  \\
\hline
 High Noise & 46 & 44  & n/a & 44  \\
\hline
 HN+PSF & 59 & 52  & 11 & 26  \\
\hline
 HN+PSF+ Foreground & 74 & 66  & 11 &  40   \\
\hline
 Low Noise+PSF & 128 & 105  & 14 & 78   \\
\hline
 LN +PSF+ Foreground & 146 & 115   & 14 & 89     \\
\hline
\hline
\end{tabular}
 \label{table:sext}
\end{table*}

\subsection{Description of Simulated Images}
\label{sec:simimage}

\begin{figure}
\begin{center}
\subfigure{\epsfig{file=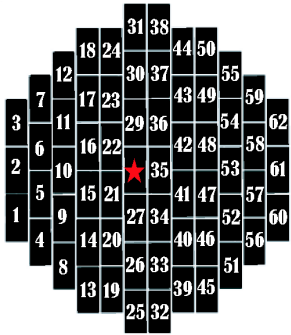,height=5cm,angle=0} } \\
\end{center}
\caption{A graphical representation of the image plane with the standard DES CCD numbering. The red star marks the 
cluster center.}
\label{fig:ccdmatch}
\end{figure}

The DES focal plane consists of 62 imaging CCDs, each with $x (y)$
dimensions of 2048 (4096) pixels.  The image plane is shown in
Fig.~\ref{fig:ccdmatch}. Since the pixel scale of the DECam is 0.27
arcseconds per pixel, the radius of the image plane is $\sim 65$
arcminutes, and the field of view is $\sim 3$ square degrees.

The truth catalog contains many pieces of information about each
object, the most useful for our purposes being:

\bi
\item (RA, DEC) and ($x,y$) pixel positions,
\item  whether the object is a star or a galaxy,
\item  redshift,
\item  $g,r,i,z,y$ magnitudes,
\item  SIS shear 
\ei
For simplicity, the images were made only in the $r$-band, but as
noted above, the truth catalog contains the magnitudes for four other
bands for each object, which are assigned according to the
prescription described in \secref{sec:simlndescrip}.  To apply the
shear from the cluster, the exact SIS value for a given location in
the image plane is convolved with the intrinsic object ellipticity
through the Shapelets shear convolution algorithm \citep{refregier}.
For simplicity, magnification from the cluster was {\it not} accounted
for in the images.

To make the cluster simulation more realistic, five of the seven
images incorporate a PSF that affects all the objects in the image.
This PSF varies in ellipticity and size across each CCD, but it has an
identical form for all 62 CCDs.  The PSF is modeled with two pieces:
the first is an isotropic Gaussian-smearing function with a slowly
linearly increasing width varying in 10 equally spaced steps, each of
about 0.02 pixels.  In total, the width varies from 2.9 pixels at the
bottom to 3.1 pixels at the top of the CCD.  The second piece of the
PSF is an anisotropic component whose total ellipticity linearly
decreases in the same 10 steps from the bottom to the top of each CCD.
In total, this ellipticity varies from 0.6\% at the bottom to 0.2\% at
the top of the CCD.

\begin{figure}
\begin{center}
\subfigure[]{\epsfig{file=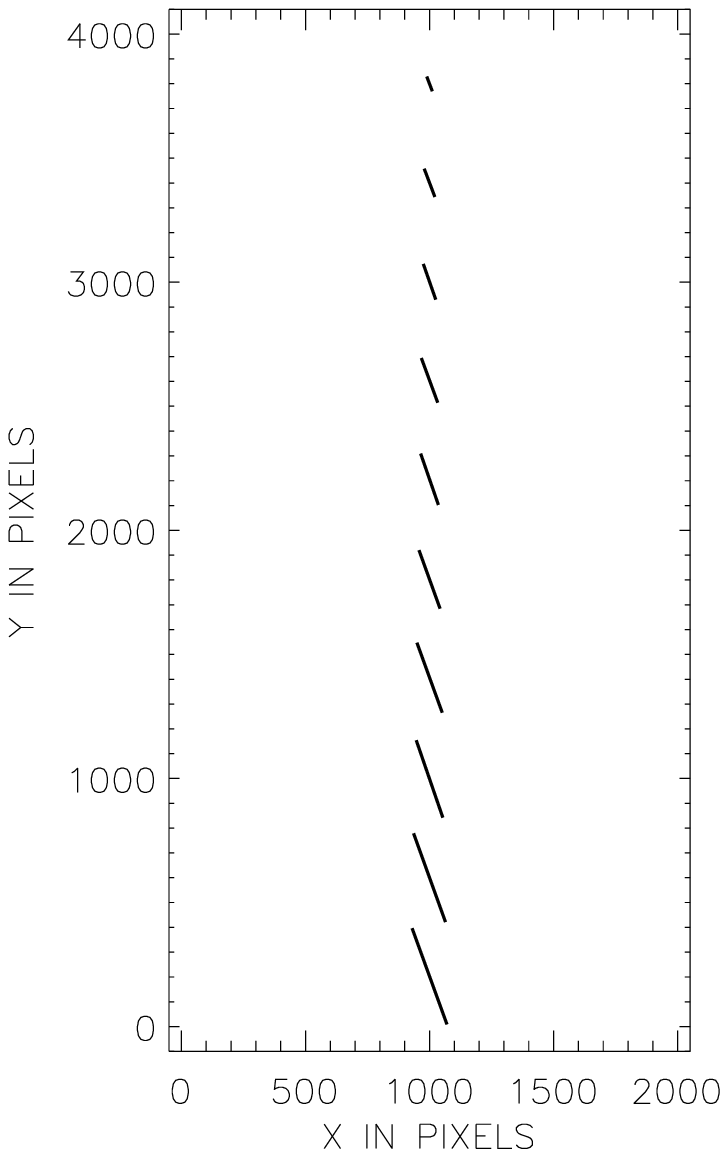,height=4.8cm,angle=0} }
\subfigure[]{\epsfig{file=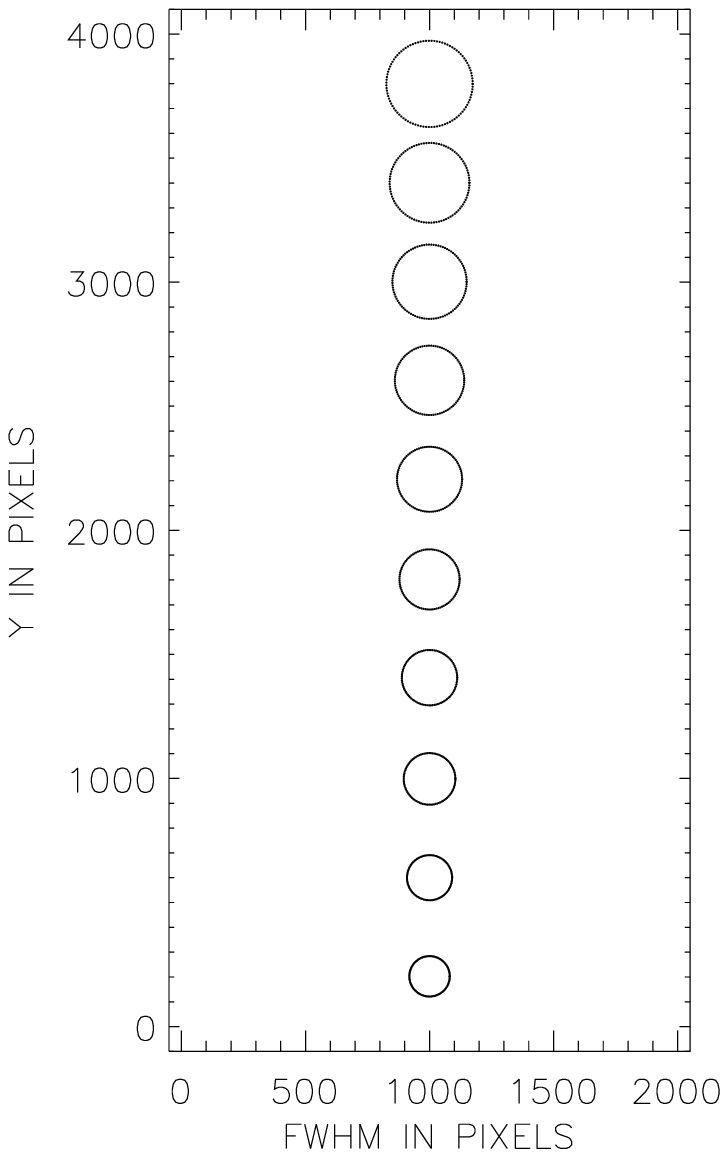,height=4.8cm,angle=0} }
\end{center}
\caption{
(a) Average of \sext\ \ellips\ in horizontal strips in each
  CCD (the PSF pattern is identical for all CCDs), with the
  inclination of the segment indicating the direction of the average,
  and the length indicating its magnitude.  (b) Average of the FWHM in
  cells in the same strips, with the radius of the circle indicating
  the scaled FWHM magnitude; specifically, with values chosen to make
  the variation visible: ${ \langle FWHM \rangle - 2.8 \over 3.2 } \times 1000$ (See
  Sec.~\ref{sec:simimage}).}
\label{fig:fwhmpsf}
\end{figure}

Fig.~\ref{fig:fwhmpsf}a shows the average of the two components of
\ellips\ for the image-selected stars in linear strips in each CCD.
The inclination of each line segments indicates the direction of the
average ellipticity, and the length indicates its magnitude. It can be
seen that the \mag\ of the \ellipy\ becomes steadily smaller towards
the top of the CCD.  In Fig.~\ref{fig:fwhmpsf}b we plot the average of
a linear function of the full width at half maximum (FWHM) for the
image-selected stars in a grid on each CCD.  Here, the radius of each
circle indicates the scaled FWHM magnitude for that strip.  It can be
seen that towards the top of the CCD the FWHM becomes steadily larger.

% Sec 3
\section{Processing through the Pipelines}
\label{sec:pipeline}

\subsection{Overview}
\label{sec:overview}

In order to process a set of either simulated or real images and
extract either a final cluster mass measurement or the SIS velocity
dispersion \sigmav, there are five fairly distinct steps
that must be executed:

\benum

\item Find objects in the image and  build a catalog of their properties.
  
\item Use these properties to determine which of the objects are stars and which are galaxies. 
  
\item Use the objects determined to be stars to measure the PSF and remove its effect 
      on the lensed galaxies.
  
\item Refine galaxy selection with cuts that attempt to pick
      only galaxies behind the cluster.
  
\item Use the shear measurements to reconstruct a shear profile of the cluster 
     and infer its mass or velocity dispersion.

\eenum

We detail these steps in the following sections.       

\subsection{Source Extraction}
\label{sec:sextractor}

As the first step in both the IMCAT and Shapelets pipeline, we execute
Step (i) of Sec.~\ref{sec:overview} and extract objects from the image
files using
% the well-known standard program 
Source Extractor (or ``SExtractor'', \citealt{bertin}).  
SExtractor runs over an image and builds a catalog of objects
% I think they *are* generally referred to as a 'sexcat'
that contains various properties. Though the two pipelines invoke
\sext\ in different ways, we use identical \param s\ in their
respective configuration files and thus obtain fully identical
SExtractor object catalogs from them with respect to the number of
objects yielded and their properties.  Our SExtractor configuration
files are available on the website.

\subsection{Star Selection}
\label{sec:starselec}

We next move to Step (ii) of Sec.~\ref{sec:overview}, separating star
and galaxy objects in the SExtractor object catalogs.  Even though the
star objects are very clear in this simulated cluster, it is useful to
verify our star selection tools, which are needed for real images.

It is very important to have a good selection of stars when
classifying objects because this determines the shape of the PSF that
is later removed.  Thus, we developed a number of tests to find the
best sample of stars for our PSF determination.  In general, stars are
selected from an image by plotting the \mag\ of the object vs. a measure
of size such as the FWHM, and then selecting out the stellar locus in
this figure, which is defined by objects that are small and fall in a
very narrow range of sizes but a wide range of \mag s.  The FWHM is a
standard output from SExtractor obtained by fitting the object with a
Gaussian profile (see Sec.~8.4.4 of \citealt{holwerda}).  The
magnitude measure we chose to use from SExtractor, the MAG\_AUTO
output, is obtained by fitting the object to an elliptical aperture
using second-order moments of its brightness (see Sec.~7.4 of
\citealt{holwerda}).

\begin{figure}
\begin{center}
%\subfigure[Sheared file ]{\label{fig:magvsfwhm3}\includegraphics[scale=.3, angle=270]{figs/FWHM_3.ps}} 
%\subfigure[Noise-applied file ]{\label{fig:magvsfwhm5}\includegraphics[scale=.3, angle=270]{figs/FWHM_5.ps}} \\
%\subfigure[PSF-applied file ]{\label{fig:magvsfwhm6}\includegraphics[scale=.3, angle=270]{figs/FWHM_14.ps}} 
\subfigure{\label{fig:magvsfwhm7}\includegraphics[scale=.3, angle=270]{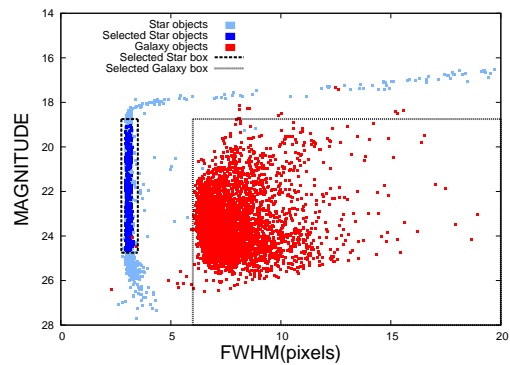}} 
\end{center}
\caption{\Mag\ ($r$-band) vs. FWHM (in pixels) for  the LN image with noise, PSF, and foreground objects.  
% Old, when we had several figs: Note that as soon as noise is added, in the Noise-applied file, the dimmest objects are no longer seen, 
% and spurious small FWHM scale noise is instead picked up by SExtractor.  
Colors indicate the star and galaxy objects -- stars in blue, and galaxies in red.  Only 10\% of the objects are shown for clarity.}
\label{fig:size_vs_magnitude}
\end{figure}

In Fig.~\ref{fig:size_vs_magnitude}, we plot the \mag\ vs. FWHM of
objects in the LN image with noise, PSF, and foreground objects.  In
blue are objects that are known from the truth catalog to be stars,
and in red are the ``truth-matched'' galaxies.  We see clearly there
the stellar locus, or ``star column'', indicated by the objects
contained in the vertical box on the left.  Specifically, to isolate
the ``good star'' objects we chose a FWHM of 2.9 to 3.3 pixels and a \mag\ from
20.3 to 26.  In the same figure we see several regions of
stars, starting from the top arm stretching to the right: these are
saturated stars which are not useful for PSF estimation.
% Why not: they bleed into neighboring pixels and become much larger than
% a normal star we use for PSF estimation.  
Moving down, we see in the vertical box in dark blue the good
stars that are selected for the PSF shape estimation, and then in the
region below this rectangular box stars that are dim and have poorly
measured shapes, which we discard.

To make the initial selection of the galaxies, we use the information
that they are generally larger than the stars in the number of pixels
they cover, and have less surface brightness per pixel.  They are thus
generally characerized by larger FWHMs and fainter apparent \mag s and
they are the red objects on the right in
Fig.~\ref{fig:size_vs_magnitude}.  The magnitude limit we have chosen
is approximately 26.5, though realistic $r$-band limits in DES will be
closer to 24.  The dimmer galaxies are in our sample mostly because
the initial catalog we are working with has an $i$-band magnitude
limit of approximately 24, but an attempt was made to make realistic
color distributions, which resulted in fainter magnitudes in the
$r$-band (see \secref{sec:simlndescrip}).  We also note in the figure
that there is a fairly sharp edge for the smallest galaxies.  This is
caused by a discrete lower cutoff in the size of the galaxies that
were included in the simulation.

% \sext\ settings which causes no smaller objects to be found.

\subsection{Galaxy Shape Estimators and PSF Removal}

Moving next to Step (iii) of Sec.~\ref{sec:overview}, we briefly
describe the estimators the two pipelines use for the shapes of the
galaxies, and the mechanisms by which they remove the PSF from the
image, beginning with some formalism that is common to both pipelines.

\subsubsection{Common Formalism for Shape Estimation }
\label{sec:common}

The initial steps are to take the first moments of the image intensity of the object
as a function of pixel position $I(x,y)$, to find its center, then the
second (quadrupole) moments which will be needed to extract the
\ellipy\ (e.g. \citealt{great}).

To form the first moments of the image $I(x,y)$, we define
\bea
\bar{x} &=& \int \, I(x,y) \, x \, \rmd x \, \rmd y, \\
\bar{y} &=& \int \, I(x,y) \, y \, \rmd x \, \rmd y
\eea
and for the quadrupole moments we define
\bea
Q_{xx} &=& \int I(x,y) \, (x-\bar{x})^2 \, \rmd x \, \rmd y, \\
Q_{xy} &=& \int I(x,y) \, (x-\bar{x})(y-\bar{y}) \, \rmd x \, \rmd y, \\
Q_{yy} &=& \int I(x,y) \, (y-\bar{y})^2 \, \rmd x \, \rmd y.
\eea

\begin{figure}
\begin{center}
\epsfig{file=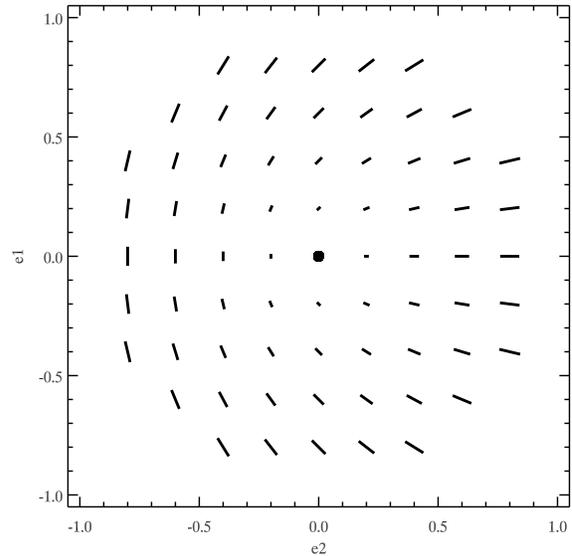,height=8cm,angle=0}  
\end{center}
\caption{Representation of object ellipticity representation with length indicating magnitude and angle indicating direction, using the parameterization in the two quantities \eone\ and \etwo\ (see eqs.\ref{eq:pzydefa}-\ref{eq:pzydefb}).
% Vs. the normal defn of $e$ and $\theta$. 
  Note how, due to its rotational symmetry, rotating any ellipse by $\pi$ yields the same  \eone\ and \etwo\ parameters. }
\label{fig:ellip} 
\end{figure}

Next, the \ellipy\ components $e_1$ and $e_2$ may be extracted by forming from the second moments the complex quantity
\be
\label{eq:equad}
e = \frac{Q_{xx} - Q_{yy}+2iQ_{xy} }
{Q_{xx}+Q_{yy}+2(Q_{xx}Q_{yy} - Q_{xy}^2)^{1/2}} \equiv  e_1 + i e_2.
\ee
For an object with constant ellipticity isophotes having major axis
$a$ and minor axis $b$ and orientation of the major axis $\theta$
with respect to the positive x-axis (see Fig.~\ref{fig:ellip}), the two
components of the \ellipy\ from \eqref{eq:equad} become

\bea
\label{eq:pzydefa}
e_1 = { a - b \over a+b} \cos 2\theta, \\
e_2 = { a - b \over a+b} \sin 2\theta.
\label{eq:pzydefb}
\eea
 
Gravitational lensing maps the unlensed image, specified by
coordinates $(x_u, y_u)$, to the lensed image $(x_l, y_l)$ using a
matrix transformation:
 
\be
\lp \begin{array}{c} x_\ell \\  y_\ell \end{array} \rp = {\cal A} 
\lp \begin{array}{c} x_u \\  y_u \end{array} \rp
\ee
where $\cal A$ is the shear matrix:
\be
{\cal A} = \lp \begin{array} {c c}  1 - g_1 & -g_2 \\
                -g_2    & 1 + g_1 \\
    \end{array} \rp
\ee
and $g_1, g_2$ are the two components of reduced shear 
\be
g = {\gamma \over 1- \kappa}
\ee
where $g$ simply becomes the normal shear $\gamma$ in the limit of the
convergence $\kappa \rar 0$, which generally holds in the weak
lensing regime for clusters, and will be assumed henceforth here.

The intrinsic \ellipy\ of a galaxy $e$ transforms under shear as 
\be
e^\ell = \frac{e^i - g}{1-g^* e^i}, 
\ee 
where $e^\ell$ is the observed
sheared (lensed) shape of the galaxy. For $|g|<1$ this quantity can be
Taylor-expanded to give 
\be 
\langle e^i \rangle \simeq g 
\ee 
to first order in $g$, where we have assumed \mbox{$\langle e_1^i
  \rangle =\langle e_2^i \rangle =0$} and the average is over a
population of galaxies.  Thus, an average of the tangential component
of the shear (the `E-mode') over a population of galaxies in an
annulus around the cluster center will yield an estimate of the shear
in that annulus, which is a fact we shall use later in our cluster
profile shear extraction in \secref{sec:results}.  The average of the
component tilted at $\pi \over 4$ relative to the tangential component
is the `B-mode' and should be exactly zero for the case of perfect
cluster shear (see e.g. Sec.~4 of \citealt{bANDs} or \citealt{bANDj}).

\subsubsection{Shapelets PSF Removal}

To execute the Shapelets PSF removal pipeline, after selecting out the
star objects, we form an initial shapelet model for each of them.  Then
we pick an optimal fixed \interpln\ order and Gaussian basis function
width common for all of these stars and form a shapelet model for each
one.  The choice of order and width that one uses will influence the
ellipticity of the shapelet objects. To determine the optimal order
and width for the decomposition, Shapelets uses a
$\chi^2$-minimization algorithm to determine the best match between
the reconstructed and original images.  After all of the stars have been
decomposed down to a shapelet model with a fixed order and width, we
determine the order of the fitting function to interpolate each of the
shapelet coefficents across the image plane and construct the PSF
model.  This constructed PSF is then used 
% via Shapelets linear algebra algorithms 
to remove the effects of the PSF shear from the galaxy objects. In
Shapelets it is possible to estimate the final shape of a galaxy in a
number of ways, in this paper we use the minimal estimator, which is a
ratio formed from three of the low order shapelet coefficients,
specifically \be \gamma_{\rm sh} \equiv {\sqrt{2} f_{22} \over f_{00}
  - f_{40} }, \ee (see e.g. eq.~(39) of \citealt{polarshlets}).

\subsubsection{IMCAT PSF Removal}

\def\psm{P^{\rm sm}_{\xi\zeta}}
\def\psh{P^{\rm sh}_{\xi\zeta}}

\def\psmstar{P^{\rm sm *}_{\xi\zeta}}
\def\pshstar{P^{\rm sh *}_{\xi\zeta}}
\def\pgamma{P^\gamma_{\xi\zeta}}
\def\pgammainv{(P^\gamma)^{-1}_{\xi\zeta}}

\def\psmstarinv{(P^{\rm sm,*})^{-1}_{\xi\zeta}}

IMCAT also uses stars to remove the PSF-induced alteration of the size
and \ellipy\ on the galaxy objects, relying on just quadrupole moment
shape measurement rather than the entire decomposition of the shape into basis
functions of an orthogonal set. It relies on the well-tested KSB
algorithm \citep{ksb} which assumes the PSF can be broken into two
parts, an anisotropic and an isotropic piece, and can be removed in
two steps.  We first remove the anisotropic part by writing the true
ellipticity of an object that would be seen without any PSF effects as

\be
e^t_\xi  =   e^{obs}_\xi - \psm p_\zeta, 
\ee
where $e^t_\xi$ is the vector representing the true \ellipy\ before
PSF effects, $e^{obs}_\xi$ is the observed \ellipy\ vector, and $\psm$ is
called the ``smear polarizability tensor''.  This quantity represents the
anisotropic part of the PSF, while $p_\zeta $ represents the anisotropy kernel
at a specific location in the image plane.  We estimate $p_\zeta $ by
saying that stars should appear perfectly circular before PSF smearing,
and thus

\be
e^{t*}_\xi  =   0 = e^{obs*}_\xi - \psmstar p^*_\zeta, 
\ee
where the asterisk superscript indicates this equation holds for stars.  Solving for $p^*_\zeta $,

\be
  p^*_\zeta   =  \psmstarinv e^{obs*}_\xi
\ee
and a bi-polynomial interpolation is used across the image plane to
estimate $\psm$ and $p_\zeta$ at the location of the galaxies. For each galaxy
\be
e^{t,g}_\xi  = e^{obs,g}_\xi - \psm \lp \psmstarinv  e^{obs*}_\xi \rp .
\ee

The next step is to remove the isotropic piece of the PSF and extract the shear via
\be
\gamma_\zeta = \pgammainv e^{t,g}_\xi
\ee
where
\be
\pgamma  =  \psh - \lp \psm \psmstarinv \pshstar \rp 
\ee
and is referred to as the ``pre-seeing shear polarizability'' in the
literature (originally by \citealt{luppino}).  Here $\psh$ is 
the ``shear polarizability tensor'' and encodes the information about
the variation of the isotropic piece of the PSF across the plane (see
\cite{hoekstra} for details of the derivation of $\pgamma$).

It has been found that the off-diagonal elements of $\psm$ and $\psh$ are close
to zero, and the diagonal elements are comparable, so each of them can
be approximated as the average of their trace multiplied by the
identity matrix(see e.g. Sec.~4.2 of \citealt{bacon}).  This approximation makes inverting the matrices
extremely rapid, has worked remarkably well in common usage, and
is the implentation of KSB that we used.

% Code from IMCAT:
% ecorr = e - emodel  for stars 
% e = e - 1/( (psm*00+psm*11) / 2 ) psm * estar
% pgamma = psh - psm (psm*^-1) psh* --> psh - (psm * (psh00 + psh11) / (psm00 + psm11) ) 
% gamma = pgamma^-1 e               --> (e0/ pgamma00 , e1/pgamma11)

\subsection{PSF Removal Test}
\label{sec:tests}

In order to test how well we are removing the PSF, we implemented a
test that we refer to as a `stars on stars' test.  Here we randomly
selected half of the stars to calculate the PSF distortion and apply
this correction to the other half of the stars.  In general, the size
of the corrected stars in such a test should become smaller and the
ellipticities of these stars should shrink substantially.

Because a linear variation in \ellipy\ and size of the stars was put
into the PSF, we expect that a linear order \interpln\ of the PSF will
be sufficient to adequately describe the PSF across each CCD.  This
is indeed what we find; numerically, after a linear order PSF
correction, taking the 200--300 stars per CCD on the files and dividing
them into two groups, we find the values in
Table~\ref{table:ellips} for the average and width of the values
of the \ellips\ of the PSF-corrected stars.  These results indicate a reduction by
between one and two orders of \mag\ in the \avg\ of each component of
the \ellipy, and a reduction by about a third in the \ellipy\ width
distribution.

\begin{table}
 \caption{Average stellar \ellips\ and their widths before and after PSF-removal stars on stars test (see \secref{sec:tests}).}
 \centering
   \begin{tabular}{@{}|l|l|l|l|l| @{}}
  \hline
  \hline
 File & $\langle \eone \rangle$ & $\langle \etwo \rangle$ & $\sigma_{\eone}$  & $\sigma_{\etwo}$  \\
\hline
\hline
Before PSF removal &  $-0.0132$ & $-0.0110$ &  0.0121 &  0.0112 \\
\hline
After PSF removal &  0.0008  &  0.0009  & 0.0095  &  0.0082 \\
\hline
\hline
\end{tabular}
 \label{table:ellips}
\end{table}

We also did this at several other orders of the PSF functional
interpolation polynomial.  As one would expect, at zeroth order
(constant stellar \ellipy\ and size assumed across whole plane), we
found that in fact the \avg\ \ellips\ were of the same order as before
correction, and the widths barely shifted.  Above order one, there was
no significant reduction in either the \avg\ \ellips\ nor widths above
that already gained by the order one PSF removal.  This confirms that
a a first order \interpln\ is sufficient to describe the PSF variation
across the CCD.  Of course, higher order \interpln\ is in general
needed for real images, which will not have such a simple PSF
structure or variation with spatial position.

% Sec 4
\section{Results}
\label{sec:results}

%%%%%%% Sec labels
%\label{fig:truthprofile}
% \label{table:truthsigmav}

%\label{fig:3.5.annularavgs}
% \label{table:sextsigmav}

%\label{fig:15profile}
% \label{table:shrsigmav}

%\label{fig:colorcut}
%\label{fig:colorsig}
%\label{fig:redsig}
%\label{fig:fwhmsig}

Having discussed our methods and tests, we now move to the results
from running both pipelines on the simulations.

\subsection{Fitting SIS Profiles}
\label{sec:fitsis}

\begin{figure}
\begin{center}
\subfigure[]{\epsfig{file=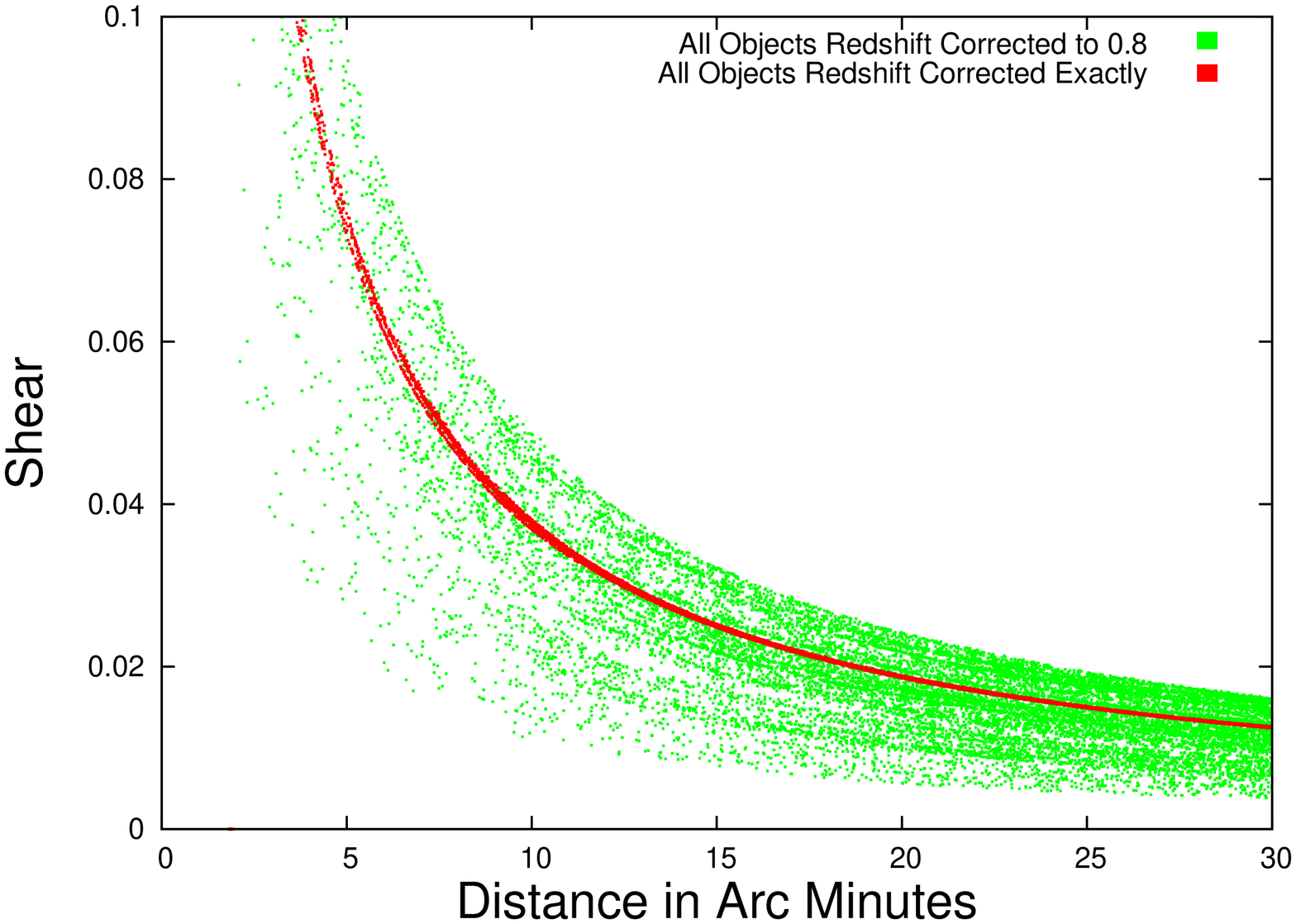, height=8cm,angle=270}}
\subfigure[]{\epsfig{file=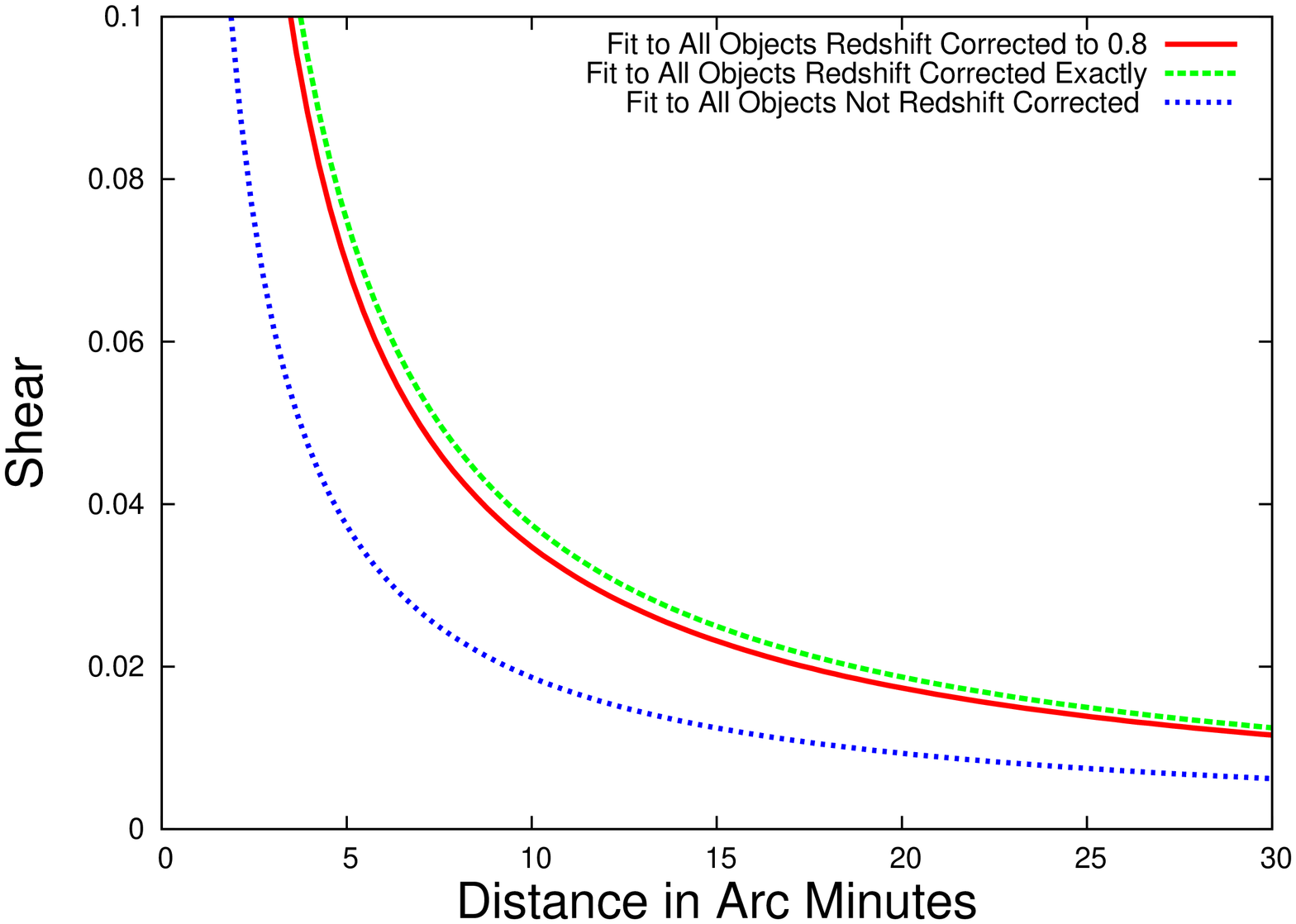, height=8cm,angle=270}}
\end{center}
\caption{Truth shear profiles for the inner half of the image plane (to see more inner profile detail)
  plotted against distance from the center of the cluster in
  arcminutes (1' = 222.2 pixels = 283 kpc for the cluster distance of
  z = 0.33 and a flat $\Lambda$CDM  geometry).  Points labelled ``Redshift
  Corrected Exactly'' are corrected point by point by the known
  redshift, points labelled ``Redshift Corrected to 0.84'' are each
  corrected assuming the single sheet approximation, i.e. that each
  galaxy lies at $z= 0.8$, which is the average of the redshifts of
  the \bkgd\ galaxies.  In (b) lines are unbinned fits to the three sets of points.}
\label{fig:truthprofile}
\end{figure}

For an SIS cluster, the critical Einstein radius for each \bkgd\
galaxy is

\be
 \theta_E = \lp {4 \pi \sigmav^2 \over c^2} \rp \beta 
\ee
where \sigmav\ is the intrinsic three-dimensional lensing cluster
galaxy velocity dispersion, and

\be
\beta \equiv {\Dls(z_l,z_s) \over D_s(z_s) },
\ee
where $\Dls(z_l,z_s)$ is the angular diameter distance from the lens
to this \bkgd\ source galaxy, and $D_s(z_s)$ is the distance from the
observer to the \bkgd\ galaxy (see e.g. \cite{bANDs},
Sec.~3.1.5).  This is the radius within which the projected average surface
mass density is exactly equal to the critical value of
\be
\Sigma_{cr} = {c^2 \over 4 \pi G } {\Dls(z_l,z_s) \over D_l(z_l) D_s(z_s)}
\ee
and within which strong lensing arcs may potentially be seen.  The shear from a
cluster on a given background galaxy depends on the Einstein radius
simply as

\be
  \gamma(\theta) = {\theta_E \over 2 \theta}.
\label{eq:shr}
\ee 
Eq.~\ref{eq:shr} is the shear of any observed galaxy, and because
of the distance ratio ${\Dls(z_l,z_s) \over D_s(z_s) }$ in the
Einstein angle, a galaxy at the same radial distance from the center
of the cluster but at a different \bkgd\ redshift is sheared
differently.  However, if the exact redshift for a specific lensed
galaxy and thus its distance ratio factor is known, we may divide
through by this and remove this variation.  This results in the
infinite redshift limit shear value, which is a unique function of the
radial distance from the cluster center

\be
  \gamma_{\infty}(\theta) = {\theta_E \over 2 \theta} \beta^{-1}.
\label{eq:thshr}
\ee 
This is exactly the shear that a galaxy at $z \rar \infty$ experiences, since
$\beta \rar 1$ as $z \rar \infty$.  As an example, for a galaxy at
$z=1$, at 1.8\arcmin (108\arcsec, or 400 pixels) from the cluster center, which
is the radius of maximal distortion put into the simulation (see
\secref{sec:clusdescrip}), the observed shear evaluates to about 0.12.
Dividing by $\beta$, this becomes the infinite redshift shear value of
about 0.21.  It then falls steadily with radius, as can be seen by the
solid red curve in Fig.~\ref{fig:truthprofile}b.

Following from eq.~\ref{eq:thshr}, if we fit the simple function of a hyperbola
\be
 f(\theta) = { \xi \over   \theta} 
\label{eq:hypfunct}
\ee
to the infinite redshift shear values, we can solve the equation 
\be
 \xi =  {\theta_E \over 2 } 
\ee
for \sigmav:

\be
 \sigmav = \sqrt{   \xi  c^2 \over 2 \pi }.
\label{eq:sigmav}
\ee

\subsection{Fits to truth shear values}

We first apply the above to the truth file shear values.  In this
case, the shear is written into the truth file as it would be
actually observed (i.e.  as eq.~\ref{eq:shr}, not eq.~\ref{eq:thshr}),
so the distance ratio factor must be corrected for each object so
that we can use eqs.~(\ref{eq:hypfunct}-\ref{eq:sigmav}).  We may do
this in two different ways and compare the results: (a) correcting
point by point by the known redshift, since we know its value for each
object in our simulated images (b) correcting approximately by assuming
each galaxy lies at the average of the redshifts of all the \bkgd\ 
galaxies, i.e., the ``single sheet approximation''.  The latter is the
approach most often taken in weak lensing work, since often 
neither the spectroscopic nor photometric redshift of each galaxy is
available.

In Fig.~\ref{fig:truthprofile} we show the results of doing the
corrections in the two ways, first in Fig.~\ref{fig:truthprofile}a by
plotting the E-mode data point results (see \secref{sec:common}), and in
Fig.~\ref{fig:truthprofile}b the fit profiles as functions of the
distance from the center of the cluster image to the edge of the image
field.  We also plot in Fig.~\ref{fig:truthprofile}b the fit profile
result for the ``redshift-uncorrected'' results
(i.e. assuming incorrectly that each galaxy is located at infinite redshift).

%The unbinned fit on the exact correction yields $ \xi = 83.36 \pm 0.13$, which gives $\sigmav
%= 1250.20 \pm 0.98 {\rm km/sec} $, within one sigma of the input value.

For all truth fit results, we use a simple least squares fitting
algorithm to the unbinned points to avoid losing information in
binning data, but will present also binned data for visual
comparisons.  Here, the unbinned fit on the exactly corrected points
yields effectively the input value $\sigmav = 1249 \kmsec $, and on all objects redshift corrected
to 0.8 yields $\sigmav = 1200 \kmsec $.  Both these results have
negligible error bars ($< 1 \kmsec$) since we are working with the
truth values of the shear, which have no shape error (see
Sec.~\ref{sec:fitsex} for a discussion on where the dominant error in
\sigmav\ arises from).  As can be seen in
Table~\ref{table:truthsigmav}, the exact correction results in a
\sigmav\ value very near the input, while the the single sheet
approximation is quite far off in this case.  However, this case has
unrealistically small error bars, and we will see that in general in
actual realistic samples, the extracted \sigmav\ value falls within
the one-sigma error bar of the input value.

Doing no redshift correction (fitting only the redshift-uncorrected
values) results in $\sigmav = 880 \kmsec $, again with small error
bars.  The lower value is obtained since we are dividing each galaxy
shape simply by unity, instead of the factor $\Dls(z_l,z_s) / D_s(z_s)
$, which is always less than unity.  Doing no redshift correction is
exactly equivalent to assuming all \bkgd\ galaxies are infinitely far
away, which is clearly not a very good assumption for lensing.

These values are summarized in Table~\ref{table:truthsigmav}.

\begin{table}
\caption{Table of the \sigmav\ values derived from the truth file with varying forms for the redshift correction for each galaxy. }
\begin{center}
\begin{tabular}{|l|l|l|l|l|}
\hline
\hline
   Redshift correction   & \sigmav\ (km/sec)   \\
   for each galaxy  &    \\
\hline
\hline
  $z =$ Infinity  & 880   \\
\hline
  $z = 0.84$  & 1200   \\
\hline
  $z =$ Exact  & 1249   \\
\hline
\hline
\end{tabular}
 \label{table:truthsigmav}
\end{center}
\end{table}

\subsection{Fitting Simulated Data}

We now do the fits to data sets that come out of the simulation images
themselves.

\subsubsection{Fitting SExtractor Files}
\label{sec:fitsex}

First, we fit the SExtractor data beginning with the Sheared and the
Original files before any noise or PSF is added, then proceed to
the file with noise, then that with PSF, and then to the final file
with noise, PSF {\it and} the foreground galaxies present.  Though
this exercise is not useful in the real world because we can never get
access to an image without a PSF or noise, it is instructive to do it
here and see just how much these complexities dilute the pure original
shear signal.  In this study we are using only information from
SExtractor without invoking either of the two lensing pipelines and
thus with no PSF correction.  That is, the shear in any given radial
bin is computed simply by using the observed ellipticity of that
galaxy as computed by SExtractor vs. doing any PSF-removal processing
through the pipelines.

In each case, we have corrected the shear for the redshift using the
single sheet approximation and setting all the \bkgd\ galaxies to be
at the \avg\ of the redshifts of all the \bkgd\ galaxies (see
Sec.~\ref{sec:fitsis} for an explanation of this, and see
\secref{sec:vary-z} for a study on results from varying the type of
redshift correction).  We find this value by doing a match of the
objects with the truth catalog, obtaining the redshift of each object,
then taking the \avg\ of these redshifts.  We present the \sigmav\ 
results in this way because we found that they were similar to those
we obtained by fitting to the exact redshift of each object, and
correspond to what is done more commonly in lensing analyses.  We now
see that with realistic errors, the extracted \sigmav\ value falls
within the one-sigma error bar of the input value.

%Note the \defn\ of \ellipy\ we have used is the linear one: $e =
%{{a-b} \over {a+b}}$, where $a$ and $b$ are the semi-major and
%semi-minor axes of the ellipse that is drawn as the best-fit
%elliptical contour around the object by SExtractor (see \secref{sec:sextractor}).

\begin{figure}
\begin{center}
\subfigure[]{\epsfig{file=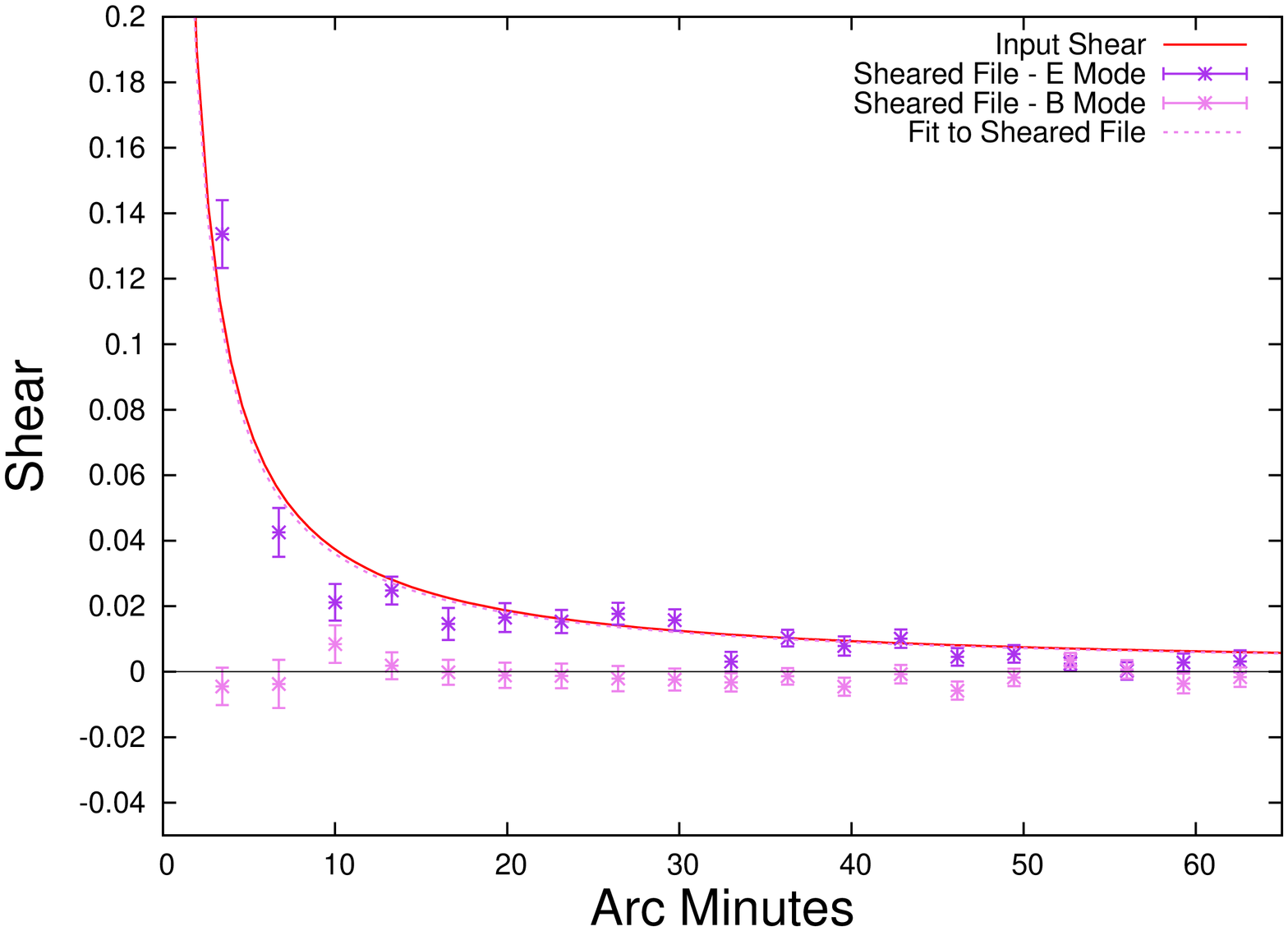, height=8cm, angle=270}} 
\subfigure[]{\epsfig{file=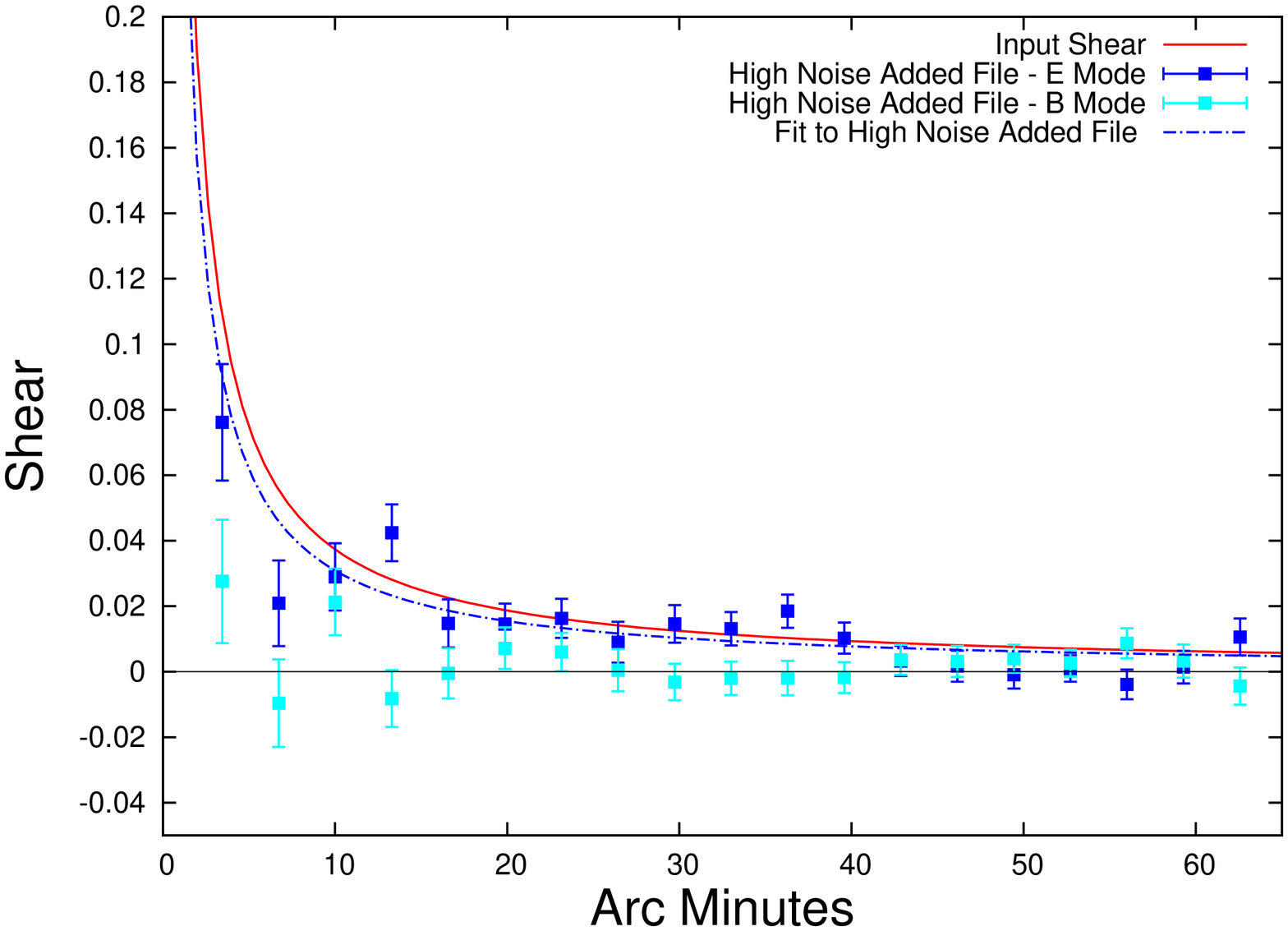, height=8cm, angle=270}} 
\subfigure[]{\epsfig{file=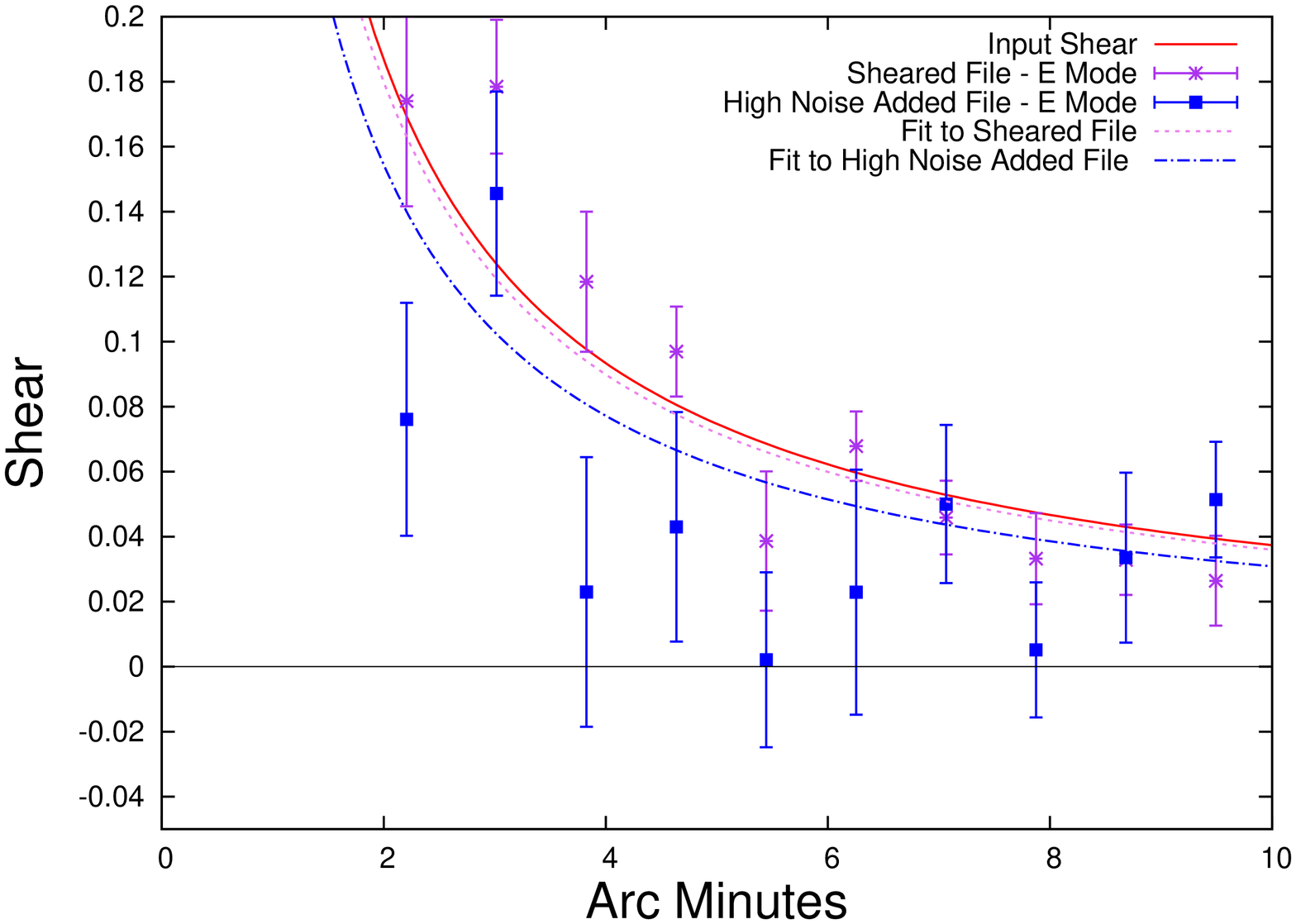, height=8cm, angle=270}} 
\end{center}
\caption{SExtractor \ellipy\ annular profile \avg s corresponding to the numerical outputs of 
  just two files from Table~\ref{table:sextsigmav}: the no noise
  sheared file (row 2 of the Table), and the sheared file with high
  noise (row 3 of the Table).  The label ``Shear'' on the y-axis is
  actually simply the mean tangential galaxy \ellipy.  The red solid line shows the
  SIS profile for the input value of \sigmav, while the points show
  the E- and B-mode profiles averaged in annular bins from the center
  of the cluster to the edge of the image.  Error bars here and in all
  binned profile plots are the rms spread in shear divided by $\sqrt{N}$, where $N$ is the
  number of galaxies in each annular bin. (a)
  Shows the result from the no noise sheared file (28.0k objects), (b) Shows the
  result from the high noise sheared file (43.6k objects), and (c) shows a zoom in of
  both, with only the E-mode data points shown. }
\label{fig:3.5.annularavgs}
\end{figure}

Table~\ref{table:sextsigmav} lists the numerical values of our
results, and Fig.~\ref{fig:3.5.annularavgs} shows the E- and B-mode
profiles averaged in 20 annular bins from the center of the cluster to
the edge of the image, compared to the input shear value.  The error
bars shown are the one-sigma standard deviation spread in \ellips\ for
all the objects in that annular bin scaled by $1 / \sqrt{N}$,
where $N$ is the number of objects in that annular bin.  Since
galaxies have an intrinsic shape dispersion of about $\sim 0.3$, this
is the dominant source of statistical error, and is reduced by $1
/ \sqrt{N}$ as more galaxies are collected per bin.  We verified
this scaling in the final fit values also by varying the number of
galaxies we used in our fits, confirming that the error in the fit
values followed a $1 / \sqrt{N}$ scaling.

We conclude from the results in Table~\ref{table:sextsigmav} that,
quite as we would expect, the \sext\ fits work best for the perfect
sheared-only file, and every additional level of complexity moves the
fit value further from the input value.  It is interesting to note
that in this case, the addition of foreground galaxies seems to have a
greater effect on the result than even the application of the PSF.

\begin{table*}
\caption{Table of the \sigmav\ values derived from the SExtractor ellipticities . The ``deviation'' is 
  the number of \(\sigma\) that the result is away from the input
  value of \sigmav = 1250 km/sec, i.e. $\sigma = { {\rm result} - {\rm
      input} \over \sigma_{\sigma_v} }$. The non-sheared file doesn't
  have an input \sigmav, so the $\sigma$ deviation is not applicable.
  Note that as the files progressively have noise, PSF, and foreground
  objects added, the deviation from the input becomes greater, and the
  error bars on the fit become larger, both of which we would expect
  since no PSF removal or foreground galaxy removal has yet been
  performed on these files.  Note foreground {\it stars} have been
  removed in all the files which contain them -- all four files with
  PSF and foreground galaxies -- by keeping only objects with FWHM$>5$
  pixels and Magnitude$>20$.  }
\begin{center}
\begin{tabular}{|l|l|l|l|l|l|}
\hline
\hline
File           &   Redshift  & \sigmav\ (km/s) & $\sigma_{\sigma_v} $ (km/s) &  Deviation  \\
\hline
\hline
Original (non-sheared) File  &   0.84  & $\sim 0$ & 123 &  n/a \\
\hline
Sheared File   &   0.84  & 1223 & 33 &  0.8 \\
\hline
High Noise File  &   0.64 & 1135 & 61 & 1.9 \\
\hline
High Noise PSF Applied File &  0.64 & 1051 & 78 & 2.5 \\
\hline
High Noise Foreground Galaxies Added File &  0.64 & 730 & 80 & 6.5 \\
\hline
Low Noise PSF Applied File &  0.84 & 954 & 26 & 11.4 \\
\hline
Low Noise Foreground Galaxies File &  0.84 & 830 & 31 & 13.7 \\
% \hline
\hline
\end{tabular}
 \label{table:sextsigmav}
\end{center}
\end{table*}

\subsubsection{Fitting PSF-removed Files}
\label{sec:psfremoved}

We now move to the core of the lensing pipelines, the results for the
files with a PSF after we have processed them through each pipeline and
removed the effect of the PSF.  In both IMCAT and Shapelets we use
first order polynomial fits to describe the PSF-variation across the
CCD, since this is the form of the PSF we initially applied to the
image.

We did tests correcting galaxies for the extracted PSF in three
different ways: (1) by extracting the PSF from a fit to the stars
in each CCD separately, as would typically be done with real data;
(2) by grouping stars from all CCDs with their relative position
on that CCD into one ``super'' CCD and extracting the PSF from a fit
to the stars in the super-CCD; (3) by extracting the PSF from a
fit to 10 `truth stars'.  These `truth stars' are extracted from an
image we constructed with one star in each of the 10 discrete
y-positions on the CCD, each affected by the PSF for that specific
y-position.

We found that the profile results changed very little whichever way we
extracted the PSF, which is reasonable in this simulation because the
PSF is replicated for each CCD, and because we found that the
SExtracted stellar properties are such a close match to the truth star
ones.
% (see \secref{sec:cluster}).  
Thus, we chose to remove the PSF in the way closest to how it would be
done in the real world, namely by extracting the PSF from a fit to the
stars in each CCD.  In future simulated images, we will more
realistically vary the PSF across the entire focal plane.  This is the
plan for Data Challenge 5 for DES, scheduled to begin in Autumn 2009.

Though we will show results from all the files with a PSF, we choose
the Low Noise file with foreground galaxies as our primary reference
file, because it has several of the major realistic features -- i.e. a
PSF, foreground stars and galaxies -- and can be processed through
both of our pipelines.  However, it does have low noise relative to
the expected DES exposures (see \secref{sec:clusdescrip}).  We expect
from the values in Table~\ref{table:sextsigmav} that this will cause
it to have a smaller error in the output \sigmav\ than the \corresp\ 
high noise version, which we examine quantitatively in
\secref{sec:vary-noise}.

%Leftover: this will make a better choice for Canonical File for our later
%studies.

After removing the PSF, we make a cut of $|(\gamma_1,\gamma_2)| < 2$
on each of the resulting two components of shear for each galaxy as is
regularly done in lensing pipelines\footnote{Technically, this is
  because removing the PSF requires a step of division by the
  polarizability.  However, the polarizability can approach zero for
  objects whose shape is badly determined, leading to the shear
  mathematically increasing to very large and unphysical values.}.
Further, though we made the initial galaxy cut to remove stars, we
next refine our selection by making a color cut of $r-i >0.7$ (see
Sec.~\ref{sec:colorcut}) to remove all the foreground galaxies.  We
choose initially to redshift-correct all objects using the single
sheet approximation (see Sec.~\ref{sec:fitsis}), since this is the
most common practice in lensing analyses, but in Sec.~\ref{sec:vary-z}
we will study how the results vary with changing this assumption. In
Table~\ref{table:shrsigmav} we show the results of this analysis on
this file, compared to doing the same fit on only the SExtractor
\ellips\ of the objects, and we see that though the error bars have
gone up by a factor of 2-3 from the SExtractor case, more
significantly, the central value from each pipeline is now within one
sigma of the input, where it was nearly 10 sigma away before.  The
error bars become larger essentially because, as part of the
PSF-removal process in each pipeline, the step of dividing by an often
small polarizability increases the intrinsic shape error of each
galaxy.  This behavior is seen in all lensing pipelines.

The results in Table~\ref{table:shrsigmav} are shown for 49k objects
that have been position-matched between the final \imcat\ and \shlets\ 
catalogs (and the truth catalog, to make the color cut to keep background
galaxies).  Another way to do this comparison is by only matching each
of the catalogs to the truth catalog so that the color cut to keep background
galaxies can be made, but not matching the objects in each pipeline
with one another, and the results for these fits are shown in the last
two lines of the same table.

\begin{figure}
\begin{center}
\subfigure[]{\epsfig{file=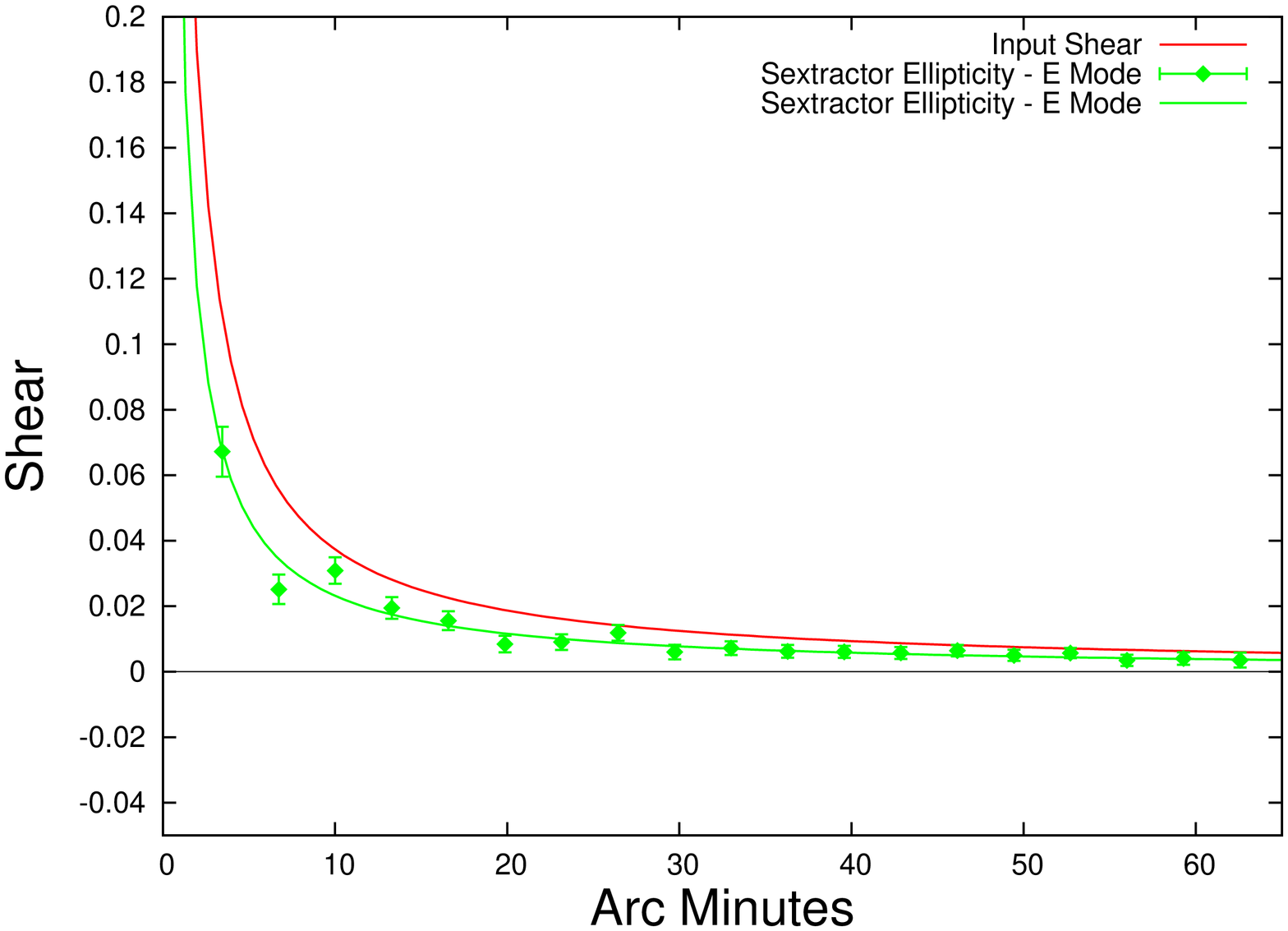,height=6cm,angle=270}}
\subfigure[]{\epsfig{file=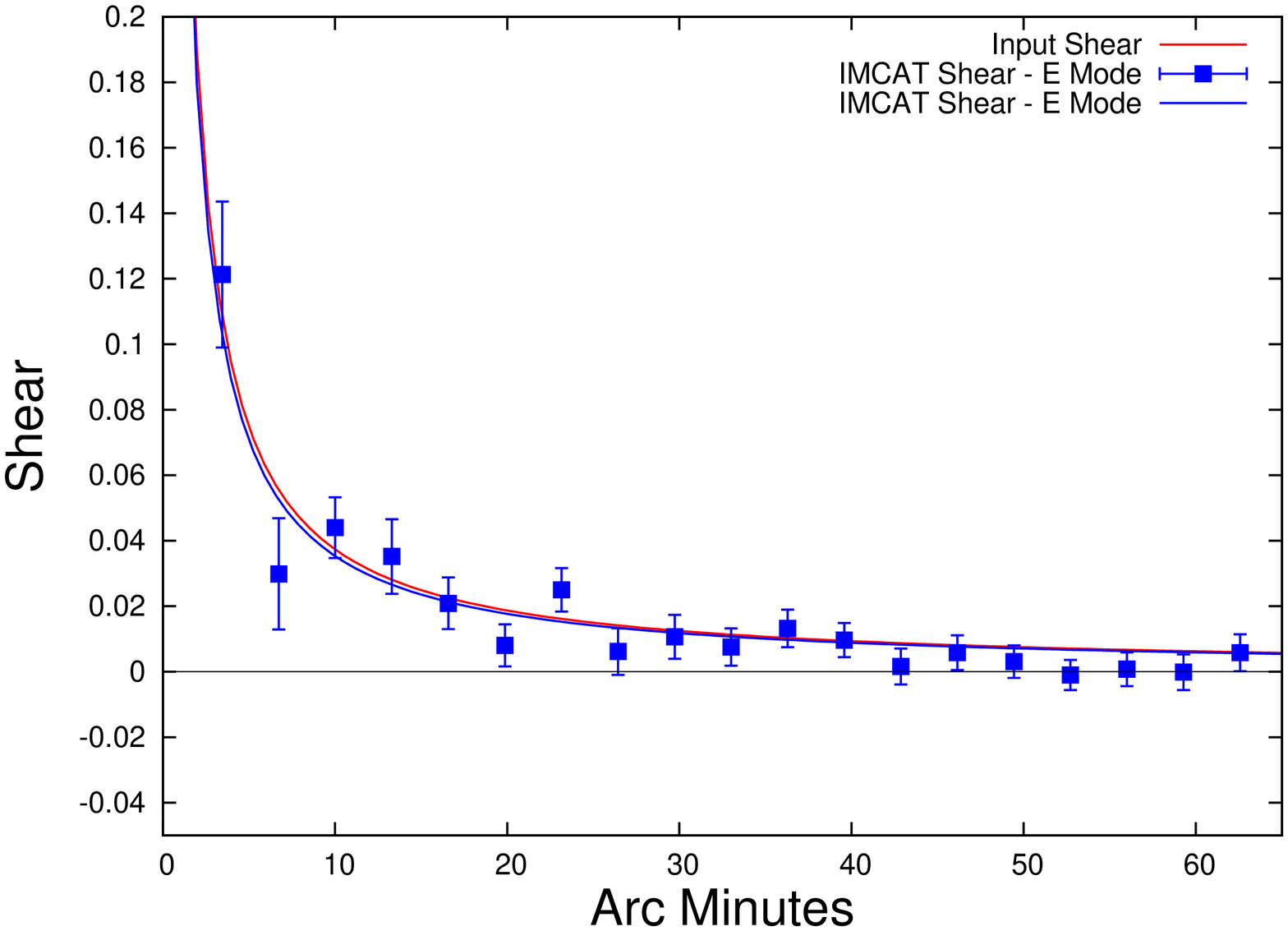,height=6cm,angle=270}}
\subfigure[]{\epsfig{file=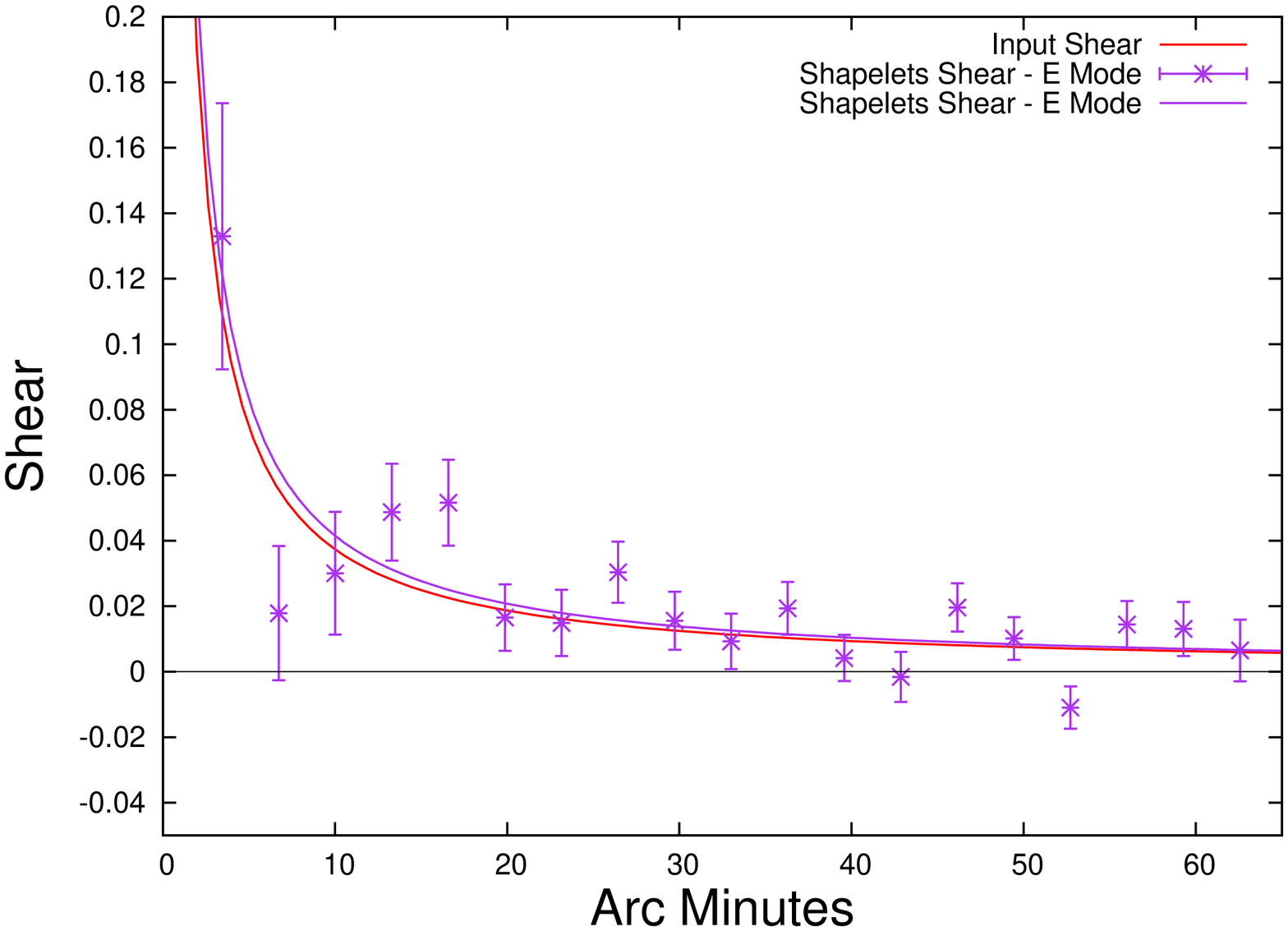,height=6cm,angle=270}}
\subfigure[]{\epsfig{file=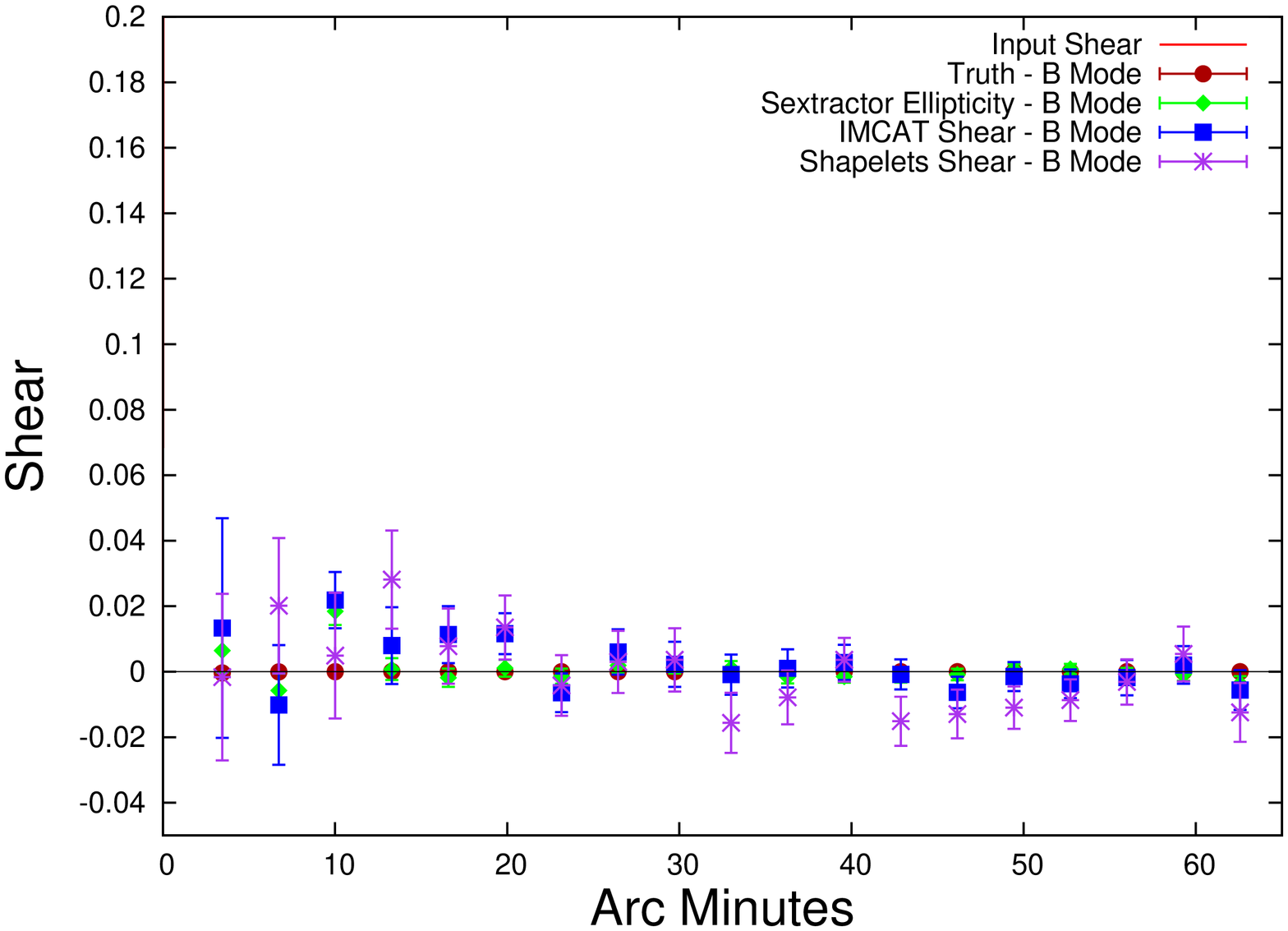,height=6cm,angle=270}}
\end{center}
\caption{Profiles for \sext\ and PSF-removed files compared to the 
  input for the entire image plane from the reference file (see
  \secref{sec:psfremoved}) -- i.e. the Low Noise file with foreground
  galaxies removed by color cut, all galaxies redshift-corrected to
  $z=0.84$ (see \secref{sec:colorcut}), and all objects
  position-matched between the IMCAT and \shlets\ files.  This results
  in a single file of 49.2k objects.  The fit profiles correspond to
  the values in \tblref{table:shrsigmav}.  The red solid line
  corresponds to the input value in all cases, and also shown are the
  binned data points and the fit profile to the unbinned points.
  Shown are the values for (a) \sext\ \ellips\ (b) PSF-removed \imcat\ 
  shear and (c) PSF-removed \shlets\ shear.}
\label{fig:15profiles}
\end{figure}

In Fig.~\ref{fig:15profiles} we show the E- and B-mode binned profiles
corresponding to the results in Table~\ref{table:shrsigmav} and compared
again to the input shear.  It is clearly seen that in the innermost
bins, the Shapelets shear is above the input, and the IMCAT shear lies
below the input, which is what leads to the \sigmav\ values of
Table~\ref{table:shrsigmav}.  The B-mode points are generally
consistent with zero across the plane and we checked that these are
well-fit to a nearly flat line in all cases.  The error bars are
generally larger for the PSF-removed points than the \sext-only points
without PSF-removal -- as mentioned above, the process of removing the
PSF increases the dispersion in shears by a large amount, and yet
the central value of the average of the corrected shapes of the
galaxies in any given bin is significantly closer to the actual input
shear.  This is indeed why the PSF is removed, though the uncertainty
in shear spread after processing through the PSF-removal pipelines is
always larger.

\begin{table*}
\caption{Values of \sigmav\ derived from the binned fits to the shear after the PSF has been 
  removed, compared to the fit to the SExtractor-only \ellips, with
  all objects from the same position-matched file (this corresponds to
  \figref{fig:15profiles}).  The deviation is as described in
  Table~\ref{table:sextsigmav}, and for the $\chi^2 / {\rm dof }$ the number of degrees of freedom is
  11 because we restrict the fit range to less than the total extent of the image plane (see text).  The central
  value from the shear output from each pipeline is within one sigma
  of the input, while it is over $11 \sigma$ away before PSF-removal.  }
\begin{center}
\begin{tabular}{|l|l|l|l|l|l|l|}
\hline
\hline
File           &   Redshift   & \sigmav\ (km/s) & $\sigma_{\sigma_v} $ (km/s) &   $\chi^2 / {\rm dof} $ & $\sigma$ Deviation \\
\hline
\hline
SExtractor Value LN PSF Applied File &  0.84 & 954 & 26 & $12.9 / 12$ & 11.4 \\
\hline
Matched IMCAT Value LN PSF Removed File &  0.84 & 1215 & 74 & $13.1 / 11$ & $-0.5$ \\
\hline
Matched Shapelets Value LN PSF Removed File &  0.84 & 1291 & 108 & $13.5/11$ & 0.4 \\
\hline
Non-matched IMCAT Value LN PSF Removed File &  0.82 & 1224 & 59 & $10.2 / 11$ & $-0.4$ \\
\hline
Non-matched Shapelets Value LN PSF Removed File &  0.80 & 1373 & 102 & $20.2/11$ & 1.2 \\
\hline
\hline
\end{tabular}
\label{table:shrsigmav}
\end{center}
\end{table*}

% Sec 5
\section{Variational Studies}
\label{sec:varstud}

Having a simulated cluster in hand for which we know all the
properties, we are able to do controlled studies from which we learn
how the extracted \sigmav\ varies with varying several types of cuts
and weightings.
 
\subsection{Variation of results with type of redshift-correction}
\label{sec:vary-z}

\begin{figure}
\begin{center}
{\includegraphics[scale=.35, angle=270]{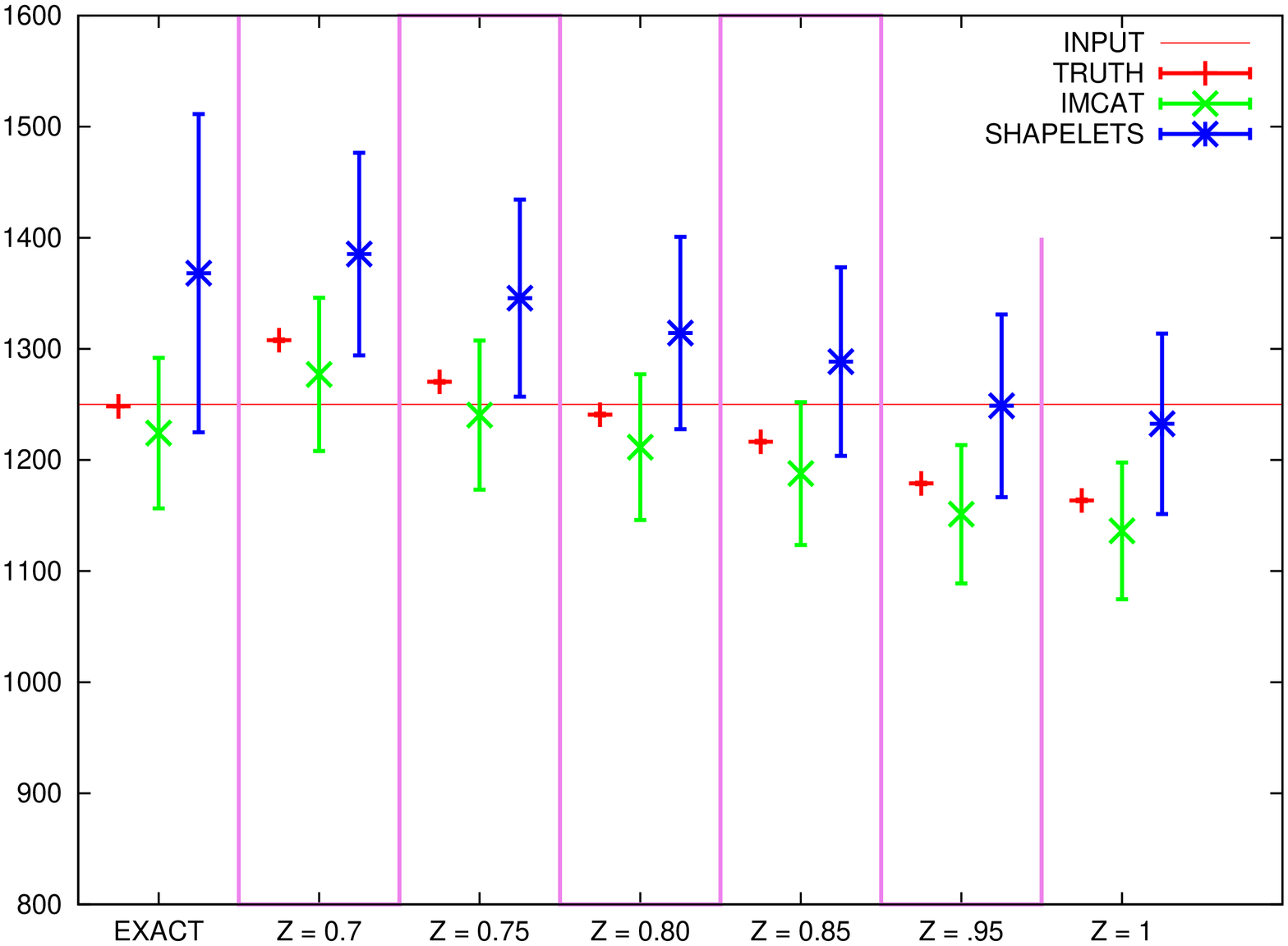}} 
\end{center}
\caption{Variation of \sigmav\  with differing ways to apply a redshift correction to the \bkgd\ objects.  
  The leftmost column shows the effect of using the exact correction
  for each object using its known redshift.  The following columns show the effect of using the \ssa\ with the objects fixed
  to varying redshifts.  The actual average redshift of all the \bkgd\ 
  galaxies is $\langle z_b^{true} \rangle = 0.84$.  The red crosses represent the
  effect of doing the correction on the truth values of the shear for
  each object, the green x's the same on the \imcat\ values of the
  shear for each object, and the red *'s the same on the \shlets\ 
  values of the shear for each object.}
\label{fig:vary-z}
\end{figure}

First, it is useful to investigate how the value we have assumed for
$\langle z_b \rangle$ of all the \bkgd\ galaxies in the \ssa\ affects
the resulting \sigmav.  In Fig.~\ref{fig:vary-z} we show the effect
of varying the type of redshift correction on the objects for
our reference file.
%Not sure this is worth saying: Interestingly, the best value for the truth is not at the
%actual $\langle z_b^{true} \rangle$, but at a value \approxy\ $\Delta z = 0.1$ below
%this.  This may be because of the distribution of the galaxies in
%redshift -- while the average occurs at $\langle z_b^{true} \rangle$, the median
%will occur lower than this.  Further, 
The variation of both the \imcat\ and \shlets\ results are
monotonic and linear downwards as $\langle z_b \rangle$ is varied
upward, with a nearly constant offset between each of results of each
of the pipelines and the truth value, and the \shlets\ value is
always higher than the \imcat\ one. We also note that the resultant
\sigmav\ is within one $\sigma$ of the input value for \ssa\ 
corrections from about $\langle z_b \rangle = \lp \langle z_b^{true}
\rangle-0.1 \rp \rar \langle z_b^{true} \rangle$.

\subsection{Galaxy Cut Variation Studies }

\subsubsection{Color Cuts }
\label{sec:colorcut}

\begin{figure}
\begin{center}
\subfigure[Color-Color Diagram CFHT]{\epsfig{file=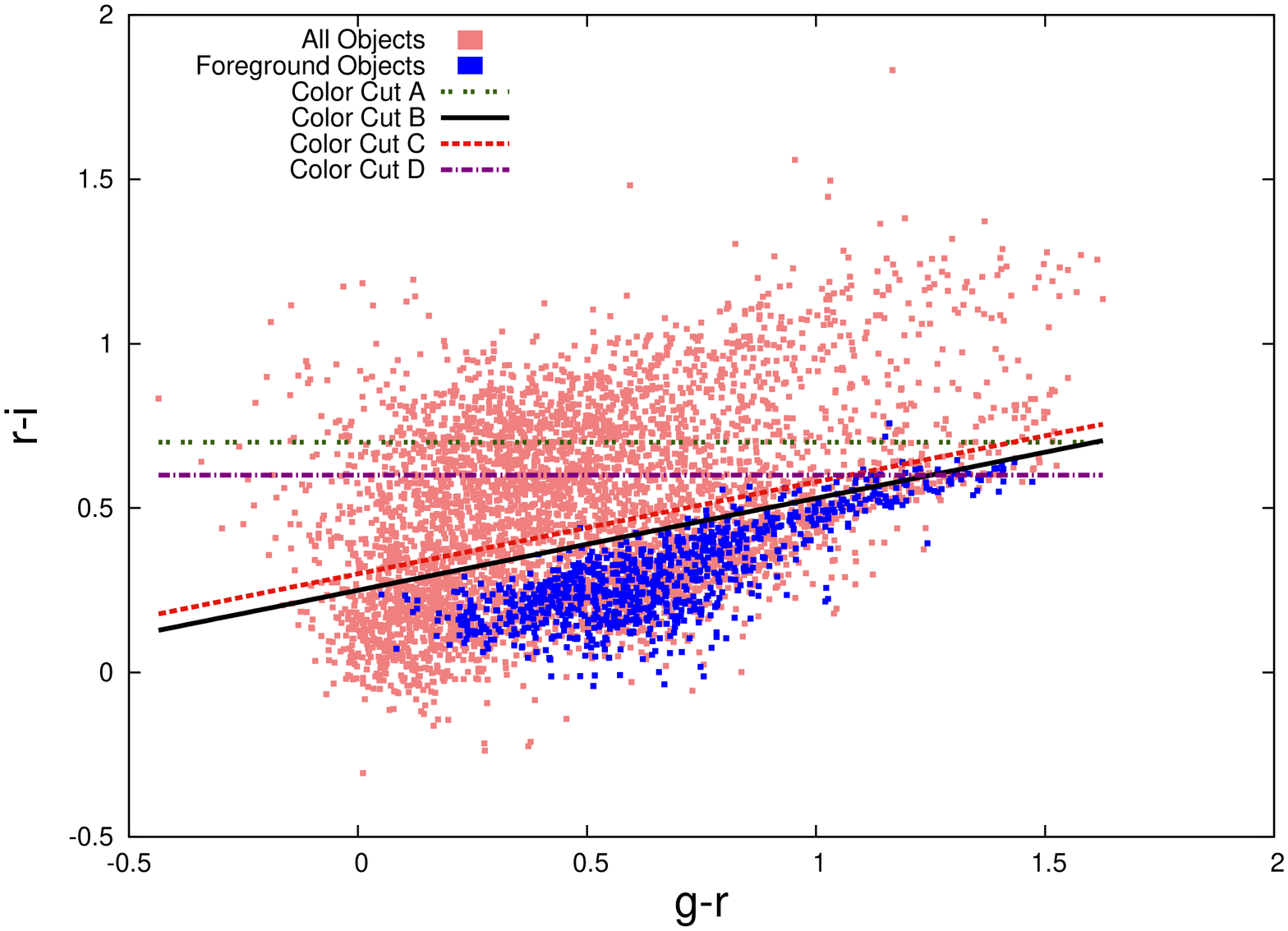,height=8cm,angle=270}} 
\subfigure[Color-Color Diagram DES]{\epsfig{file=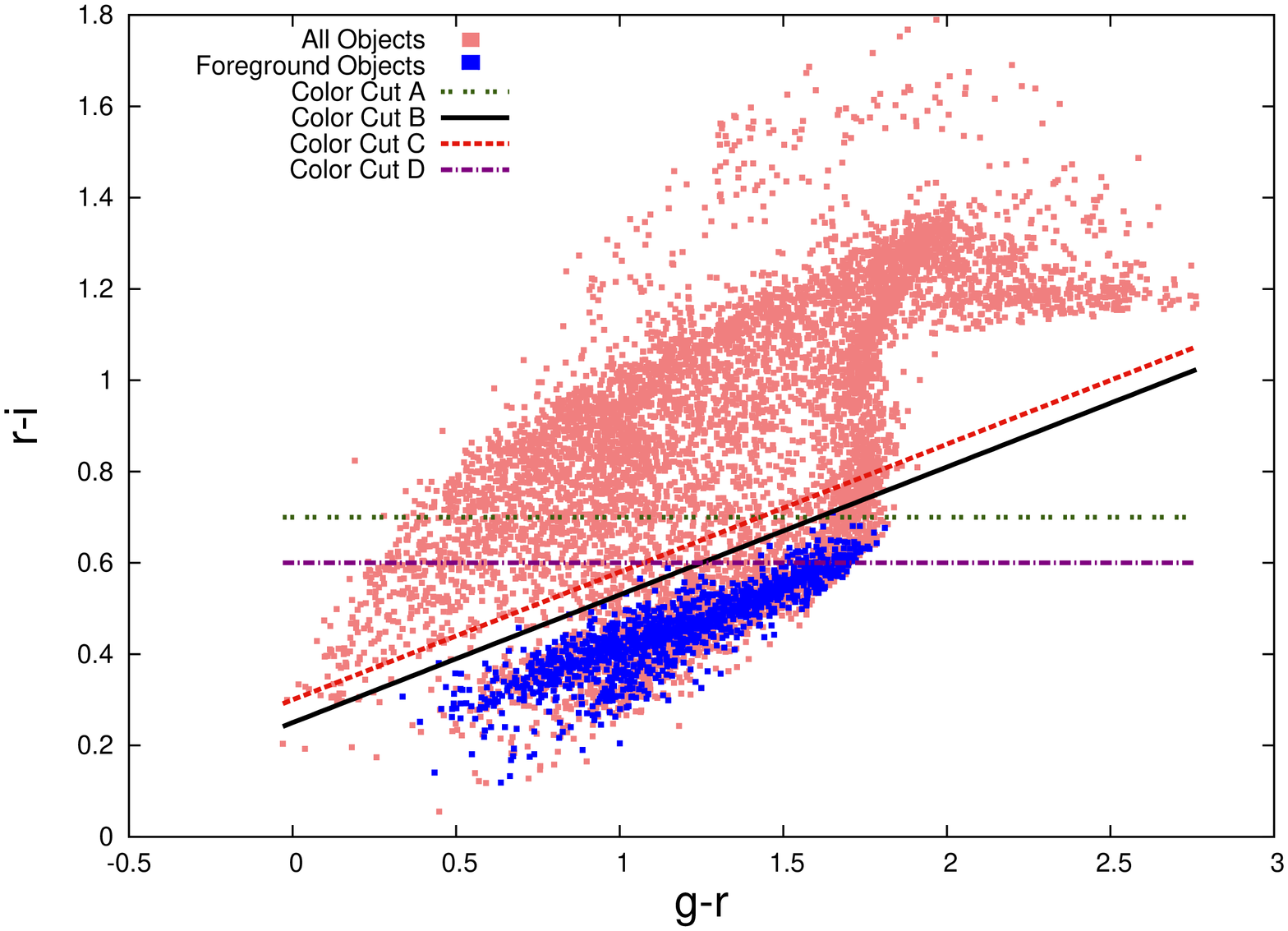,height=8cm,angle=270}} 
\end{center}
\caption{Color-Color Plots for the Low Noise file with foreground galaxies.  For CFHT,  color and redshift information is from observations, while for our simulated image, 
  the color and redshift information is obtained from a match to the
  truth files, for which such information is available (see
  Sec.~\ref{sec:simimage}) .  ``Color Cut A'' selects the objects for
  which $r-i$ is greater than 0.7.  ``Color Cut B'' selects the objects
  for which $r-i$ is greater than $(g-r) \times 0.28 +0.25$.  ``Color Cut C''
  selects the objects for which $r-i$ is greater than $(g-r) \times 0.28 +0.3$.
  ``Color Cut D'' selects the objects for which $r-i$ is greater than
  0.6. Note that though the Color-Color diagrams for CFHT and DES look
  fairly different, the same color cuts do seem to have similar
  effects in how much foreground contamination is removed for both of
  them.}
\label{fig:colorcut}
\end{figure}

\begin{figure}
\begin{center}
{\includegraphics[scale=.35, angle=270]{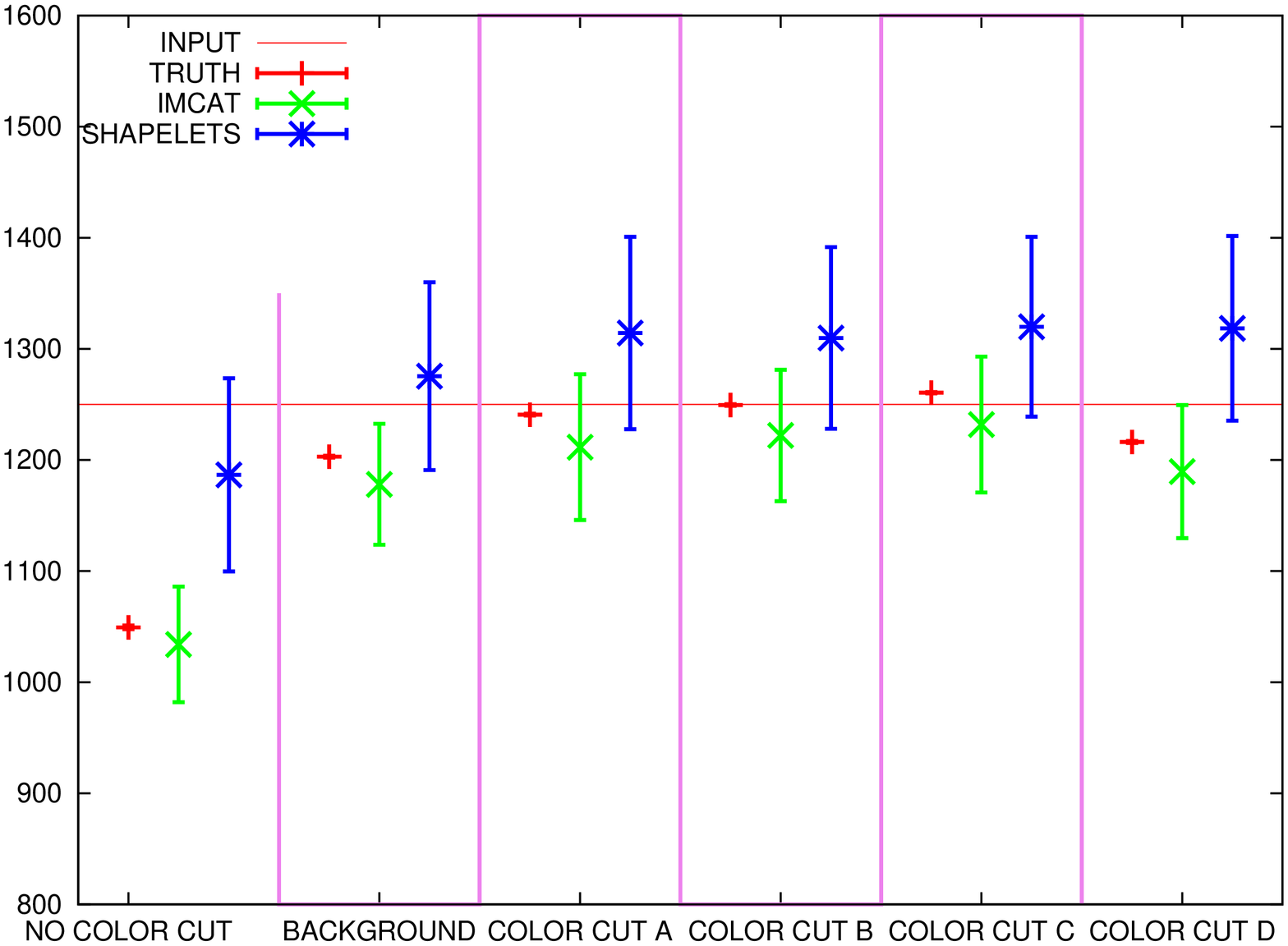}} 
\end{center}
\caption{Variation of \sigmav\  with varying color cuts.  Symbols are as in \figref{fig:vary-z}, and cut labelling is as in \figref{fig:colorcut}. All objects are corrected to the \ssa\ value of  $\langle z_b \rangle = 0.84$.  In the column labelled ``\Bkgd'' we have used the truth information on the redshift of each object to select all the true \bkgd\ galaxies.  }
\label{fig:vary-colorcut}
\end{figure}

In order to most effectively remove the foreground and cluster galaxies 
% Normal defn of foreground gals: the galaxies which are in front of the cluster or are cluster members, and
% thus not sheared -- 
when one has no redshift information on the
galaxies, one generally makes a color cut on objects, as foreground
and \bkgd\ galaxies are located in different loci of color space. By
examining the distribution of the simulated objects, for which precise
redshifts and magnitudes in various colors are known, it is possible
to determine a color cut which will remove most of the foreground
galaxies. For the reference file, we chose the simple $r-i > 0.7$ cut
shown in Fig.~\ref{fig:colorcut} (Cut A), and we describe in the same
figure three other ways one might make this color cut.  In
Fig.~\ref{fig:vary-colorcut}, we show the effect varying the color cut
has on the determination of \sigmav.  We see from this figure that the
various possible ways of making color cuts have no dramatic effect on
the \sigmav\ values, except for Color Cut D for which the fit values
for the truth and \imcat\ deviate the most from the others.  As in the
previous section, the \shlets\ results are consistently larger and
both \imcat\ and \shlets\ generally track the truth value.

%Maybe: Cut D contains clearly
%the most amount of foreground contamination, though it has the closest
%result to the \bkgd-only galaxies as well, so we see that there is
%some kind of compensation for doing the \ssa\ at $\langle z_b^{true} \rangle$ by
%adding in some amount of foreground galaxies (see \secref{sec:vary-z}).

%Maybw: thus we see
%that purity is more significant than efficiency in extracting the
%profile; i.e. making more sure when there is a tradeoff that there is
%minimal foreground contamination than that we have captured the
%maximum number of \bkgd\ galaxies.

\subsubsection{Size Cuts } 

\begin{figure}
\begin{center}
{\includegraphics[scale=.35, angle=270]{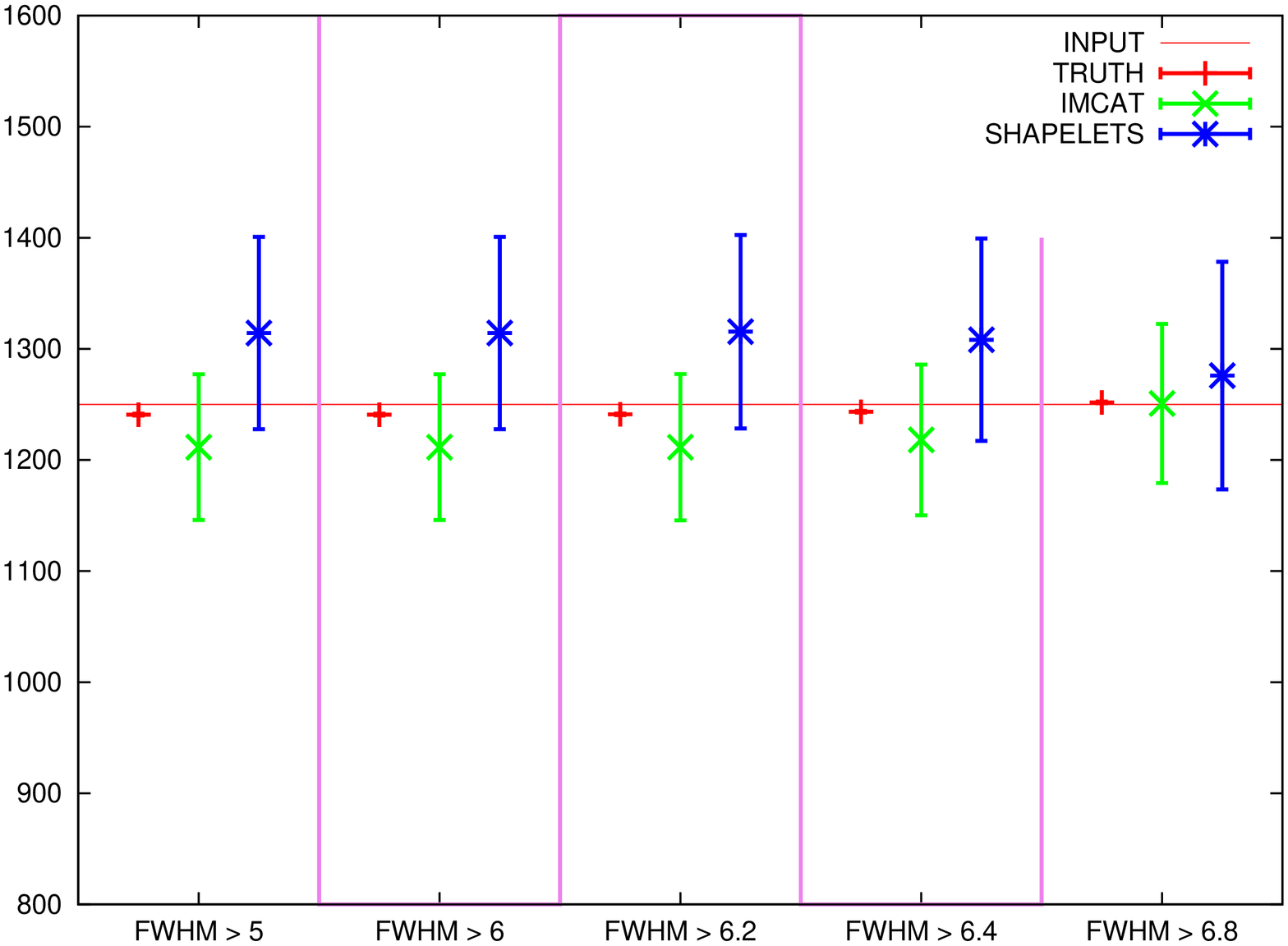}} 
\end{center}
\caption{Variation of \sigmav\ for various FWHM Cuts.  Symbols are as in \figref{fig:vary-z}. All objects are corrected to the \ssa\ value of  $\langle z_b \rangle = 0.84$. The cuts range from ``no cut'' in the leftmost panel (i.e. FWHM$>5.0$ pixels, as is already done for the initial galaxy selection) to  FWHM$>6.0,6.2,6.4,6.8$ pixels in the following four panels. }
\label{fig:vary-fwhm}
\end{figure}

Various cuts on size are used to attempt to eliminate the error due to
poor ellipticity and shear estimation. A larger object should in
general have less error in its ellipticity and shear estimation. The
effect of various cuts on the FWHM are detailed in
Fig.~\ref{fig:vary-fwhm}.  We see mild variation in \sigmav\ as a
function of FWHM cut, with both the \shlets\ and \imcat\ \sigmav\ 
tending slowly toward the input as smaller objects are cut out.  This 
indicates that the better measurement of the shape of larger objects
yields somewhat better final results.  However, there is a tradeoff in
that the error bars are also seen to grow steadily as the statistics
are reduced (by about 49k to 37k galaxies from the mildest to
strictest cut, a factor of about 25\% fewer objects).  Since
magnification is not included in our simulation, we of course cannot
quantify from these images yet further effects on the shear profile
such as the effect of faint and small background galaxies becoming
larger and magnified enough to be seen behind the cluster, also referred to as ``lensing bias''
(\citealt{schmidt}, \citealt{schmidt2}).

% No longer in here: \subsubsection{Cluster Center Variation }
% Detailed studies showed that variations within 140 pixels (about 180 kpc) have little effect, 
% and this is far beyond how badly we would expect to misestimate the center of the cluster.

% No longer in here: \subsubsection{Fit Limit Variation }
% Varying the fit limits has very little effect, as long as the lower limit is above 400 pixels away from the center 
% (the point at which the shear is first applied to objects).

\subsubsection{Noise Variation }
\label{sec:vary-noise}

\begin{figure}
\begin{center}
{\epsfig{file=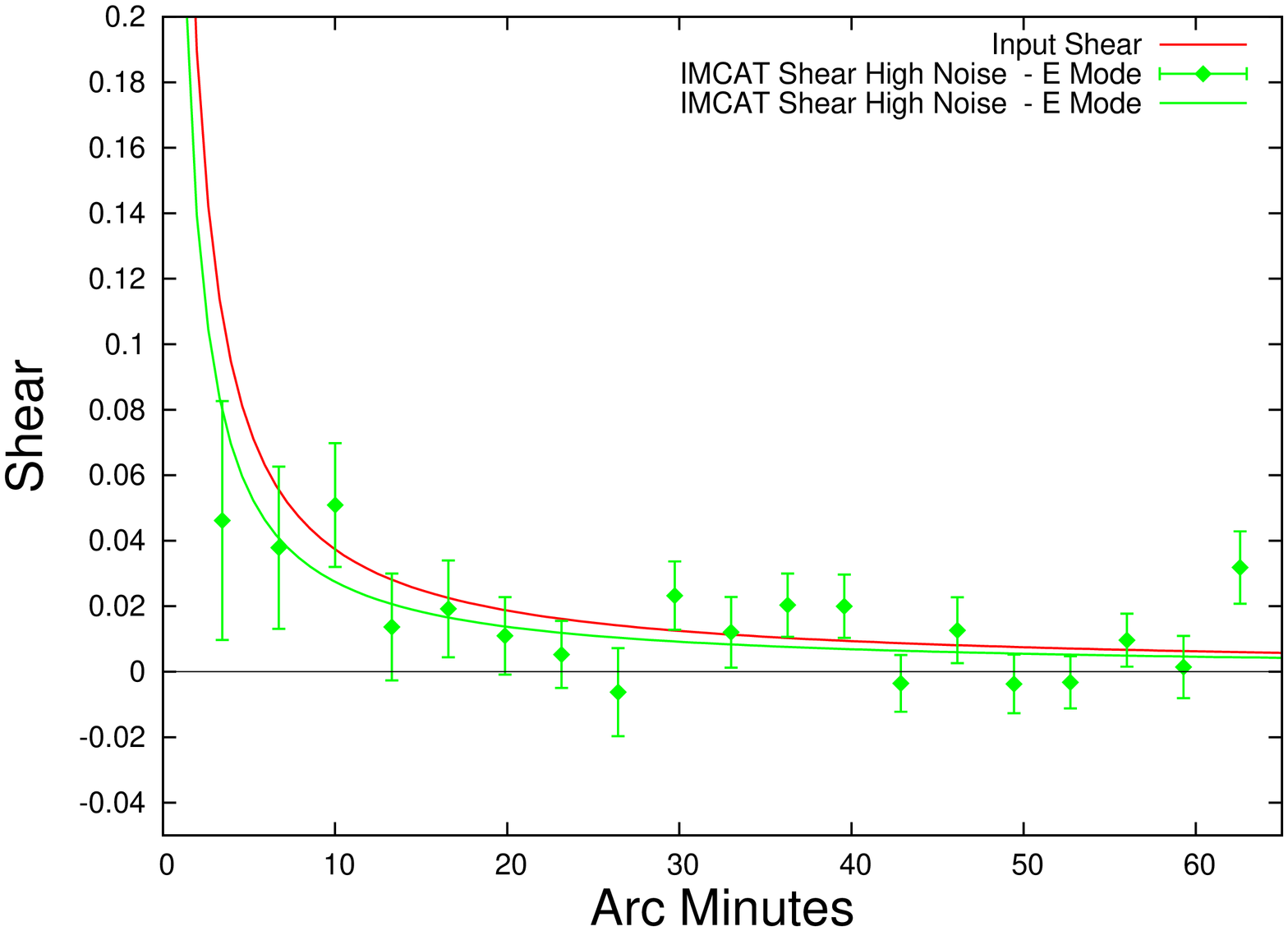,height=8cm,angle=270}} 
{\epsfig{file=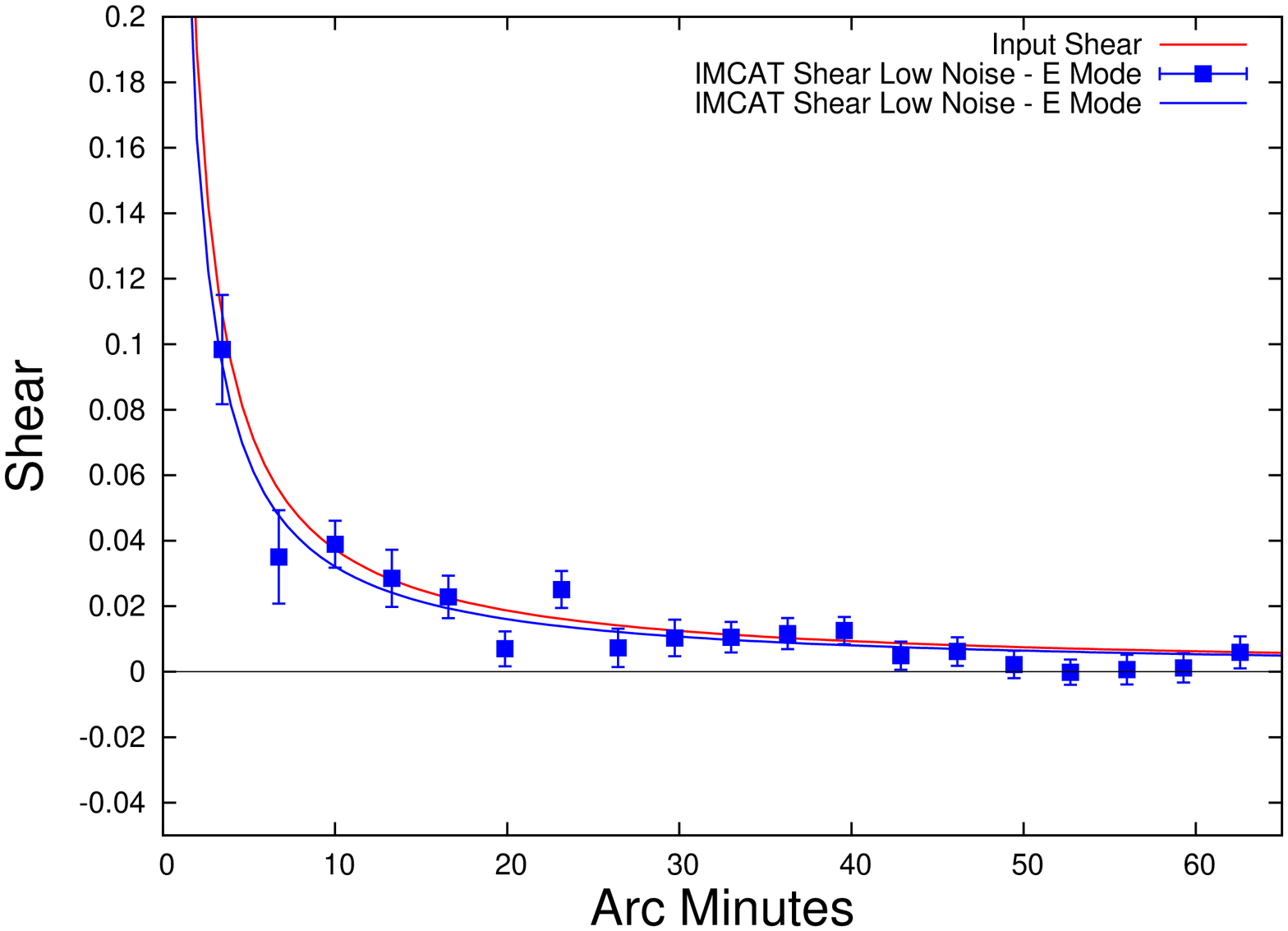,height=8cm,angle=270}} 
\end{center}
\caption{Fit profiles and binned points for two levels of noise.  
  All objects for the (a) High Noise file are corrected to the \ssa\
  value of $\langle z_b \rangle = 0.64$ (23.1k objects), All objects for the (b) Low
  Noise file are corrected to the value of $\langle z_b \rangle = 0.84$ (49k
  objects). The $\chisq \over \dof$ is 13.2/12 for the LN file, and 8.2/12 for the
  HN one. Note the larger error bars on the HN points (see text in
  \secref{sec:vary-noise}).}
\label{fig:vary-noise}
\end{figure}

Shapelets was not able to work on the High Noise files\footnote{This
  is regularly seen, and occurs essentially because \imcat\ needs only
  a few parameters to describe the shape of the ellipse that
  circumscribes an object, while \shlets\ aims to extract the total
  shape information of an object into a coefficient basis, and will
  fail more often in low S/N situations where the edges are less
  well-defined.}, so the results in \figref{fig:vary-noise} are for
\imcat\ only.  The Low Noise values have smaller error bars than the High
Noise values, and this is only partly accounted for by the smaller number of
objects in the HN images.  In fact, the LN file has about 49k objects,
and the HN one 23k, and the error bars on the extracted \sigmav\ are
74 km/sec vs.  135 km/sec, where if we scaled just by statistics on
the LN error we would obtain 108 km/sec.  Thus, the HN error is a
factor of about 25\% higher than just statistical scaling, indicating
that the higher noise has led to a wider spread in the shear estimates
of the objects.

From this we see that higher noise is harmful to profile extraction
not only for suppressing statistics, but intrinsically in worsening
shape estimation of objects.

\subsection{Bootstrap Results }
\label{sec:bootstrap}

As a simple test to verify that the differences in the central values
for \sigmav\ were significant for the cut variations discussed above,
we did a bootstrap test for two different cuts (varying the galaxy
size cutoff) in one pipeline (\imcat). To do this, we used one example
original file to first make 1000 bootstrap samples (choosing 21k out
of 21k galaxies randomly, so that some galaxies were not chosen, and
others were picked more than once).  We then applied the two differing
cuts on each sample, creating a pair of files for each of the samples.
Next, we extracted the two \sigmav\ values for each sample pair, and
took the difference of these values.  Finally, we compared the width of
the distribution of this difference to that of the \sigmav\ 
distribution of the original 1000 file bootstrap sample.  We found
that the former was smaller by more than a factor of five, indicating
a strong correlation between each sample pair -- i.e.  the cuts were
affecting each bootstrap sample in the same direction.  This indicates
that the observed differences are in fact significant in this case.

% Taken out: In \figref{fig:bootstrap} we show the results of doing the fits on the
%two different bootstrap samples.  The difference in width between
%\figref{fig:bootstrap}a and the much narrower \figref{fig:bootstrap}b
%can be seen quite clearly here.

We further probed this by evaluating a standard correlation
coefficient test on the samples, and found a 98\% correlation in the
results in the case where the cuts were made on the same exact samples
as above.  This fell to an 11\% correlation in bootstraps where the
pair of cuts were not made on random files in the ensemble.  This
further indicates that the differences between the central values in
the samples after cuts are indeed significant as we have been
assuming previously.

%\begin{figure}
%\begin{center}
%{\includegraphics[scale=.35, angle=270]{figs/fitvals.twocutfiles.ps}} 
%{\includegraphics[scale=.35, angle=270]{figs/fitvals.twocutfiles.difs.ps}} 
%{\includegraphics[scale=.35, angle=270]{figs/fitvals.twocutfiles.ratio.ps}} 
%\end{center}
%\caption{Histograms of SIS fit values with two cuts: (a) 100 fits for each histogram (b) differences in the 100 pairs of fit values.
%  The horizontal scales for the two histograms are different, and the
%  second has a much narrower width than those in the first plot (see
%  \secref{sec:bootstrap}).  }
%\label{fig:bootstrap}
%\end{figure}

% Sec 6
\section{Conclusion}
\label{sec:conclusions}

We have run two independent weak lensing analysis pipelines on a suite
of realistic simulated galaxy cluster images created in the framework
of the simulation tools developed by the Dark Energy Survey.  This
simulation contains noise, foreground objects, and a PSF which varies
both in FWHM and ellipticity. Both pipelines give results that are
within the one-sigma error bars of the input value using the same
processing that would be applied to real data.  Thus, our analysis of
this DES cluster simulation demonstrates that both the \imcat\ and
\shlets\ pipelines are able to reproduce the input singular isothermal
sphere \sigmav\ in a realistic simulation, suggesting that current
weak lensing tools are accurate enough for extracting the shear
profile of massive clusters in upcoming large Stage III \citep{detf}
photometric surveys such as DES.

These new simulations are quite a bit more realistic than the STEP
simulation images to test pipelines for cluster weak lensing profile
extraction accuracy, since with real data one must in general remove a
PSF that varies in size and ellipticity across the focal plane. With
this simulation it was possible to also test more realistic galaxy
selection cuts such as those made on color and size, as well as vary
the fit limits and star selection, and demonstrate that the pipelines
were able to handle these different choices robustly.  It seems as
though IMCAT can handle images with a high level of noise more
robustly than Shapelets, although both pipelines performed better on
images with a low level of noise.  While this indicates that Shapelets
may not be indicated for future rapid cadence surveys, we do point out
we have only worked thus far with our implementation and configuration
of the method and code.

Using both of these independent pipelines and potentially others in
parallel on the same future cluster data samples will provide an
internal crosscheck which will give confidence on shear estimates and
masses extracted from cluster weak lensing analyses.  We will follow
this study with future analyses on suites of multiple weakly lensed
cluster images with differing noise levels with different shear
profiles (such as NFW) to do statistical tests on the pipelines. The
dominant errors are from noise and galaxy statistics, thus it is
important to have a large enough galaxy sample in an image with as low
as possible noise properties in order to get the optimal shear
profile.

We are making the suite of images used in this analysis publically
available for others to analyze at {\bf
  http://ccapp.osu.edu/DEScluster}.

\section*{Acknowledgments}

The simulated images used in this study are based on the DES mock
galaxy catalog created by M.~Busha and R.~Wechsler, and we fully express
our appreciation to them for being able to use these catalogs.

We would like further to very much thank D.~Applegate, J.~Beacom,
J.~Berge, G.~Bernstein, S.~Bridle, J.~Cohn, D.~Clowe, T.~Eifler,
G.~Evrard, J.~Frieman, S.~Habib, B.~Jain, M.~Jarvis, C.~Kochanek,
K.~Kuehn, S.~Kazantzidis, A.~Leauthaud, R.J.~Massey, T.~McKay,
R.~Nakajima, C.~Orban, B.~Ragazzone, B.~Rowe, E.~Sheldon, A.~Slosar,
G.~Steigman, M.~White, H.~Yan, and A.~Zentner for many helpful
discussions, pointers and readings.

\label{lastpage}

\end{document}